\documentclass[useAMS,usenatbib,a4paper]{mn2e}

\usepackage{graphicx, epsfig}
\usepackage{float}
\usepackage{subfig}
\usepackage{natbib}
\usepackage{cancel}
\usepackage{mathrsfs}
\usepackage {amsbsy,amssymb,amsmath}
\usepackage{hyperref}

\newcommand{\ee}{\end{equation}}
\newcommand{\bea}{\begin{eqnarray}}
\newcommand{\eea}{\end{eqnarray}}
\newcommand{\bk}{{\bf k}}
\newcommand{\bl}{{\pmb \ell}}

\newcommand{\mpa}{m_\parallel}
\newcommand{\mpe}{m_\perp}
\newcommand{\ipe}{i_\perp}
\newcommand{\ipa}{i_\parallel}
\newcommand{\jpe}{j_\perp}
\newcommand{\jpa}{j_\parallel}
\newcommand{\Npe}{N_\perp} 
\newcommand{\Npa}{N_\parallel}

\newcommand{\kpa}{k_\parallel}

\newcommand{\bell}{{\pmb\ell}}
\newcommand{\bL}{\mathbf{L}}
\newcommand{\bx}{\mathbf{x}}
\newcommand{\phis}{\phi^\star}

\def\simlt{\lower.5ex\hbox{$\; \buildrel < \over \sim \;$}}
\def\simgt{\lower.5ex\hbox{$\; \buildrel > \over \sim \;$}}
\def\simpt{\lower.5ex\hbox{$\; \buildrel \propto \over \sim \;$}}

\newcommand{\rhob}{\ensuremath{\bar{\rho}}}

\begin{document}

\title[21\,cm lensing]
{Weak lensing of Cosmological  21\,cm radiation}
\author[ Pourtsidou  \& Metcalf ]{ A. Pourtsidou$^{1,2}$ \& R. Benton Metcalf$^{1}$  \\ 
 $^{1}$Dipartimento di Fisica e Astronomia, Universit\'{a} di Bologna, viale B. Pichat 6/2  , 40127, Bologna, Italy \\
$^{2}$Institute of Cosmology \& Gravitation, University of Portsmouth, Burnaby Road, Portsmouth, PO1 3FX, United Kingdom
}
%

\title{Gravitational Lensing of Cosmological 21~cm Emission}
\maketitle

\begin{abstract}
We investigate the feasibility of measuring weak gravitational lensing using 21~cm intensity mapping with special emphasis on the performance of the planned Square Kilometer Array (SKA).  We find that the current design for SKA\_Mid should be able to measure the evolution of the lensing power spectrum at $z \sim 2-3$ using this technique.  This will be a probe of the expansion history of the universe and gravity at a unique range in redshift.
 The signal-to-noise is found to be highly dependent on evolution of the neutral hydrogen fraction in the universe with a higher HI density resulting in stronger signal.  With realistic models for this, SKA Phase 1 should be capable of measuring the lensing power spectrum and its evolution.  The signal-to-noise's dependance on the area and diameter of the array is quantified.  
We further demonstrate the applications of this technique by applying it to two specific coupled dark energy models that would be difficult to observationally distinguish without information from this range of redshift. We also investigate measuring the lensing signal with 21~cm emission from the Epoch of Reionization (EoR) using SKA and find that it is unlikely to constrain cosmological parameters because of the small survey size, but could provide a map of the dark matter within a small region of the sky.
\end{abstract}


\section{Introduction} 

21cm cosmology is a new and exciting area of research with a great deal of potential. Future radio telescopes
like the SKA\footnote{www.skatelescope.org}  (Square Kilometer Array) will give us access to previously unexplored epochs of the Universe, such as the Epoch of Reionization (EoR), the end of the Dark Ages and the beginning of the Cosmic Dawn \citep{Dewdney:2013}.

Another important development in the field has been the HI intensity mapping technique in which the distribution of galaxies can be measured without detecting individual galaxies. Instead, the 21cm emission is treated as a continuous (unresolved) background, much like the Cosmic Microwave Background (CMB), but extended in the frequency (redshift) dimension. Several groups are planning to use this technique to measure the Baryon Acoustic Oscillations (BAO) scale at redshifts of order unity \citep{Chang:2008,Chang:2010,Seo:2009fq, Masui:2010, Ansari2012,Battye:2012tg,Chen2012,Pober2013,Smoot2014,Bull:2014rha}, an important probe of dark energy. 

By combining results from different kinds of surveys (CMB, galaxies, 21cm), which probe complimentary redshift ranges, we will be able to span a very large observational volume and more precisely investigate the acceleration of the Universe. Of particular relevance to this paper is the possibility of extending the cosmological probes to redshifts between those accessible with galaxy redshift surveys and the high redshift of the CMB. 
By studying the evolution of expansion and structure formation across a wide range of redshift we can investigate whether dark energy or modified gravity effects are present at a higher redshift than the standard cosmological model predicts \citep{Cope06,Clifton12}. 

The 21cm radiation provides an excellent source for gravitational lensing studies.
Earlier work \citep{Zahn:2005ap,Metcalf:2009} had shown that if the EoR is at redshift 
$z \sim 8$ or later, a large radio array such as SKA could measure the lensing convergence power spectrum and constrain the standard cosmological parameters --- however, these studies assumed a much larger survey area than is currently planned for the largest such experiment, SKA\_Low (see Section~\ref{Sec:EoR} for details). The authors extended the
Fourier-space quadratic estimator technique, which was first developed by 
\citep{Hu:2001tn} for CMB lensing  observations to three dimensional
observables, i.e. the $21$ cm intensity field $I(\theta,z)$.  These
studies did not consider 21~cm observations from redshifts after
reionization when the average HI density in the universe is much smaller.

In \cite{PourtsidouMetcalf:2014} we extended the aforementioned studies to redshifts 
after reionization, but before those probed by galaxy surveys in the 
visible bands and showed that lensing can be measured using the HI intensity mapping technique.  
Here we develop this concept further. In Section~\ref{sec:prelim} we present a preliminary overview of our study, which spans two distinct regimes (EoR and $z \sim 2-5$), and give the formal form of the lensing estimator and lensing reconstruction noise. In Section~\ref{Sec:EoR} we study how well the current SKA design will be able to map the lensing convergence at typical EoR redshifts. In Section \ref{sec:LensIM} we investigate weak lensing intensity mapping with a SKA-like interferometer array at redshifts $z \sim 2-3$. Different models for the HI mass function and different array configurations are considered. We concentrate on the ability of different SKA phases to probe the lensing signal and calculate the corresponding signal-to-noise predictions. In Section~\ref{Sec:coupledDE} we show that these measurements can be used to differentiate between some specific and  novel interacting dark energy models. We conclude in Section~\ref{sec:conc}. Throughout the paper we adopt a flat $\Lambda$CDM Universe with the PLANCK cosmological parameter values \citep{Ade:2013}, unless otherwise stated. \\

\section{Preliminaries}
\label{sec:prelim}

We will investigate the lensing of 21~cm emission in two distinct regimes.  The first is from the EoR and the second is from $z \sim 2-5$.  In the first case, the neutral hydrogen fraction is high and HI gas is not restricted to individual galaxies.  In this case the HI distribution will be approximated as a Gaussian random field that roughly follows the distribution of matter.  More sophisticated models are under investigation, but will not be considered here.  The biggest source of noise for 21 cm observations at these high redshifts comes from foreground contamination.  SKA\_Low is being planned to probe this regime.  Current plans call for it to survey a $5^\circ  \times 5^\circ$ sky area with a frequency range of $40-250 \, {\rm MHz}$ corresponding to redshifts $z \sim 5-35$ \citep{Dewdney:2013}.  

At lower redshift the HI fraction is much lower and essentially all the HI gas is in discrete galaxies.  In this regime we treat the HI distribution as consisting of discrete sources that are clustered according to the standard CDM paradigm, but also exhibit random Poisson fluctuations.  This is the standard way of modeling the distribution of galaxies in redshift surveys.  We find that this Poisson or shot noise contribution to the galaxy clustering is important for measuring lensing from 21 cm after (and perhaps during) reionization.  The evolution in the HI fraction as a function of redshift is a matter of some debate and speculation and has an important effect on the expected lensing signal to noise.  We will address this uncertainty in Section~\ref{sec:LensIM} by adopting several different models.

The quadratic lensing estimator for the CMB or a single redshift slice of the 21~cm emission works by essentially measuring differences in the power spectrum in different regions of the sky which results in correlations in Fourier (or spherical harmonic) modes that would not exist otherwise. For the CMB the local power-spectrum becomes anisotropic because of shearing and the acoustic peaks are slightly shifted to larger (smaller) angular scales by gravitational magnification (demagnification).

The effect of lensing can be divided into a shearing which can make the local power spectrum anisotropic and magnification which isotropically scales the local power spectrum.  
It can be shown that an isotropic magnification cannot be measured (with a quadratic estimator) if the power-spectrum is scale free ($C_\ell \propto \ell^{-2}$).  It can also be shown the the shear cannot be measured if the power-spectrum is constant (see \cite{Bucher:2012} for a nice demonstration of this).  As we will see, the projected matter power-spectrum in the CDM model is approximately a constant at large scales (small $\ell$) and $C_\ell \propto \ell^{-2}$ at small scales so to the extent that the HI distribution follows the dark matter distribution, the quadratic estimator picks up the magnification on large scales and the shear on small scales and some combinations in between.   

After reionization the remaining HI resides in discrete galaxies.  The galaxies are clustered in the same way as matter on large scales (modulo a bias factor), but on small scales their discreteness enters into the power-spectrum as shot noise or Poisson noise.  The power-spectrum of this component is flat (for equal luminosity sources, $C_\ell = 1/\eta$ where $\eta$ is the number density on the sky) so only the magnification can be measured.  Magnifying (demagnifying) a region of the sky reduces (increases) the number density of sources.  The result is that the power-spectrum will have features on the scale of the lensing.

Unlike the CMB, the 21~cm emission will be observable at many different redshifts.  The 3D lensing estimator effectively stacks redshift slices.
Since the lensing is coherent for different slices at the same angular position, but the 21~cm emission is statistically independent, the residual effect can be attributed to lensing.  This will be made more explicit in the next section.

\subsection{The quadratic lensing estimator}

Here we will give a more explicit description of the lensing estimator and noise.
The advantage of 21cm lensing is that one is able to combine
information from multiple redshift slices, and that the 21cm signal extends to far smaller angular scales than CMB fluctuations \citep{Loeb}, which are suppressed by Silk damping at $\ell > 3000$.  The intensity fluctuations,  $I(\theta,z)$,  are expressed in discrete Fourier space in the radial direction owing to the finite band width (wave vector $k_\parallel =
\frac{2\pi}{{\cal L}}j$, where ${\cal L}$ is the depth of the observed
volume) and in continuous Fourier space  perpendicular to the line of
sight (wave vector $\mathbf{k_\perp}=\mathbf{l}/{\cal D}$ where ${\cal D}$ the angular diameter distance to the source redshift and $\mathbf{l}$ is the dual of the angular coordinate on the sky).  See Appendix~\ref{app:intens-mapp-lens-continious} for the connection between estimators in discrete and continuous Fourier space.  Considering modes with different $j$ independent, an optimal
estimator can be found by combining the individual estimators for
different $j$ modes without mixing them.

The quadratic estimator for the lensing potential $\Psi$ is thus of the form 
\begin{equation}
\hat{\Psi}(\bL)=\sum_j \sum_{\bl}g(\bl,\bL,j) \; \tilde{I}_{\bell,j}\tilde{I}^*_{\bell-\bL,j},
\end{equation} where $\tilde{I}_{\bell,j}$ is the discrete Fourier transform of the lensed intensity field $\tilde{I}({ \pmb{\theta}},x_\parallel)$. 
The form of the filter $g(\bl,\bL,j)$ and the expected noise in the estimator depends on whether the source is discrete or continuous.  In appendices~\ref{app:intens-mapp-lens-continious} through \ref{sec:lens-estim-clust} we derive the cases that are used in the following sections. 

The mean observed brightness temperature at redshift $z$ due to the average HI density is
$\bar{T}(z)=180 \, \Omega_{\rm HI}(z) \, h\frac{(1+z)^2}{H(z)/H_0} \; {\rm mK} $
 \citep{Battye:2012tg}.
The angular power spectrum of the displacement field (the displacement of points on the sky caused by gravitational lensing) for sources at a redshift $z_s$ can be calculated with 
\begin{equation}
\label{eq:CL}
C_L^{\delta \theta \delta \theta} = \frac{9 \Omega^2_m H^3_0}{L(L+1) c^3} \int_0^{z_s}dz \, P(k=L/{\cal D}(z),z) \frac{[W(z)]^2}{a^2E(z)},
\end{equation} \citep{ Kaiser:1992}  in the weak field limit where $W(z)=({\cal D}(z_s)-{\cal D}(z))/{\cal D}(z_s)$, $E(z)=H(z)/H_0$ and $P(k)$ is the power spectrum of matter . The convergence power spectrum is simply $C^{\kappa \kappa}_L = (L^2/4) \, C_L^{\delta \theta \delta \theta}$. Note that the Limber approximation \citep{Limber} has been used in deriving Eq.~(\ref{eq:CL}). This approximation is quite accurate even for large scales $\ell \sim 10$, as the resulting formula involves an integral over the line of sight which effectively averages out the effect of the approximation, see  \citep{Schmidt:2008mb} for a detailed discussion.

The lensing reconstruction noise includes the thermal noise of the array which -assuming a uniform telescope distribution- is calculated using the formula 
\begin{equation}
\label{eq:CellN}
C^{\rm N}_\ell = \frac{(2\pi)^3 T^2_{\rm sys}}{B t_{\rm obs} f^2_{\rm cover} \ell_{\rm max}(\nu)^2} \, ,
\end{equation} where $T_{\rm sys}$ is the system temperature,
$B$ is the chosen frequency window, $t_{\rm obs}$ the total observation time, $D_{\rm tel}$ the diameter (maximum baseline) of the core array, $\ell_{\rm max}(\lambda)=2\pi D_{\rm tel}/\lambda$ is the highest multipole that can be measured by the array at frequency $\nu$ (wavelength $\lambda$), and $f_{\rm cover}$ is the total collecting area of the core array $A_{\rm coll}$ divided by $\pi(D_{\rm tel}/2)^2$ \citep{Furlanetto2006}. 

 Assuming that the temperature (i.e. the 21cm emissivity) is Gaussian distributed we find that the lensing reconstruction noise $N(L)$ is (see \citep{Zahn:2005ap} and Appendix~\ref{app:intens-mapp-lens-continious})
\begin{equation}
N(L,\nu) =  \left[\sum_{j=1}^{j_{\rm max}} \frac{1}{L^4}\int \frac{d^2\ell}{(2\pi)^2}  \frac{[\mathbf{l} \cdot \mathbf{L} C_{\ell,j}+\mathbf{L} \cdot (\mathbf{L}-\mathbf{l})
C_{|\ell-L|,j}]^2}{2 C^{\rm tot}_{\ell,j}C^{\rm tot}_{|\mathbf{l}-\mathbf{L}|,j}}\right]^{-1},
\end{equation}
where $C^{\rm tot}_{\ell,j}=C_{\ell,j}+C^{\rm N}_\ell$ and $C_{\ell,j}=[\bar{T}(z)]^2P_{\ell,j}$ with  
$P_{\ell,j}$ the underlying dark matter power spectrum. One can also consider stacking multiple bands $\nu$ and then the noise is reduced as $N_L = 1/\displaystyle\sum_{\nu} [N(L,\nu)]^{-1}$. The Gaussian approximation formalism will be used in the next Section to investigate the possibility of measuring the lensing signal with 21 cm emission from the Epoch of Reionization.

\section{Lensing studies using 21cm radiation from the Epoch of Reionization}
\label{Sec:EoR}

Observing the EoR through 21~cm has been one of the primary science goals for several large radio telescope projects including the SKA.  The collecting area and resolution of  LOFAR\footnote{www.lofar.org} or MWA\footnote{www.mwatelescope.org} are unlikely to be high enough to observe gravitational lensing as we will see.  For this reason we concentrate on the SKA\_Low instrument here which is the part of SKA designed for observing the EoR and the cosmic dawn.

The current SKA plans call for a 25 square degree survey\footnote{ Here we note that suggestions for a three-tier SKA\_Low survey have been 
recently put forward (Leon Koopmans \& Jonathan Pritchard, private communication). These include an additional 200 square degree survey in 10 $\times$ 100 hrs 
as well as a 2000 square degree survey in 100 $\times$ 10 hrs. Note that these numbers assume 
1 beam only, and a multi-beam request ($N_{\rm beam}=5$) has also
been put forward. This would further increase the sky area surveyed in a given amount of time and would greatly enhance 
the SKA\_Low science output, adding the exciting possibility of measuring the cosmological parameters (see \citep{McQuinn:2005hk} and \citep{Metcalf:2009}).} with SKA\_Low \citep{Dewdney:2013}. The convergence power spectrum measured in this size field will be dominated by sample variance.  This will make it difficult to measure the cosmological parameters through linear growth in the matter power spectrum at a competitive level with these observations even if the signal-to-noise in the lensing map is very high. This is not true of the SKA\_Mid at lower redshift where the survey area will be much larger --- see Section~\ref{sec:LensIM}. 

However, it might be possible to map the lensing convergence within the 25 square degree EoR survey area with high fidelity.  This would allow us to actually ``see'' the distribution of dark matter in a typical region of the sky, something that is only possible with galaxy lensing around very atypical, large galaxy clusters. 

For the EoR, the convergence (or, equivalently, the displacement field)
estimator and the corresponding lensing reconstruction noise are
calculated assuming that the temperature (brightness) distribution is
Gaussian, and consider the reionization fraction $f_{\rm HI}=1$ until the Universe is rapidly and uniformly reionized at redshift $z_s$.   Before reionization this is probably a very good approximation.  
Reionization is not expected to be homogeneous and may extend over a significant redshift range
 so for some period there will be ionized 
regions that grow and intersect until they fill almost all of space when reionization is complete.  Because the contrast in the brightness temperature is larger during this period, detecting and mapping the 21~cm emission will be easier, but it will change the power spectrum and  increase the fourth order moments of the brightness temperature \citep{Metcalf:2009}.  It is not yet clear how this will affect the detectability of lensing. To assess this will require numerical simulations which we plan to present in a future publication.  
Note that a semi-analytic model used to describe the patchy regime in \citep{Zahn:2005ap} showed an
increase in noise due to a decrease of the fluctuation level of the 21cm signal on the smallest resolved scales and the contribution of the connected four-point function which acts as a sample variance term correlating different $k$-modes. Of particular importance is the nonlinear structure effect during the EoR. We will be able to account for these effects once more details about the reionization process are known, but under the assumption that these features appear in higher resolution than that of our considered interferometers there is no need for the connected four-point contribution to the variance of the quadratic lensing estimator to be taken into account \citep{Kovetz:2013}.

\begin{figure}
\centerline{
\includegraphics[width=0.45\textwidth]{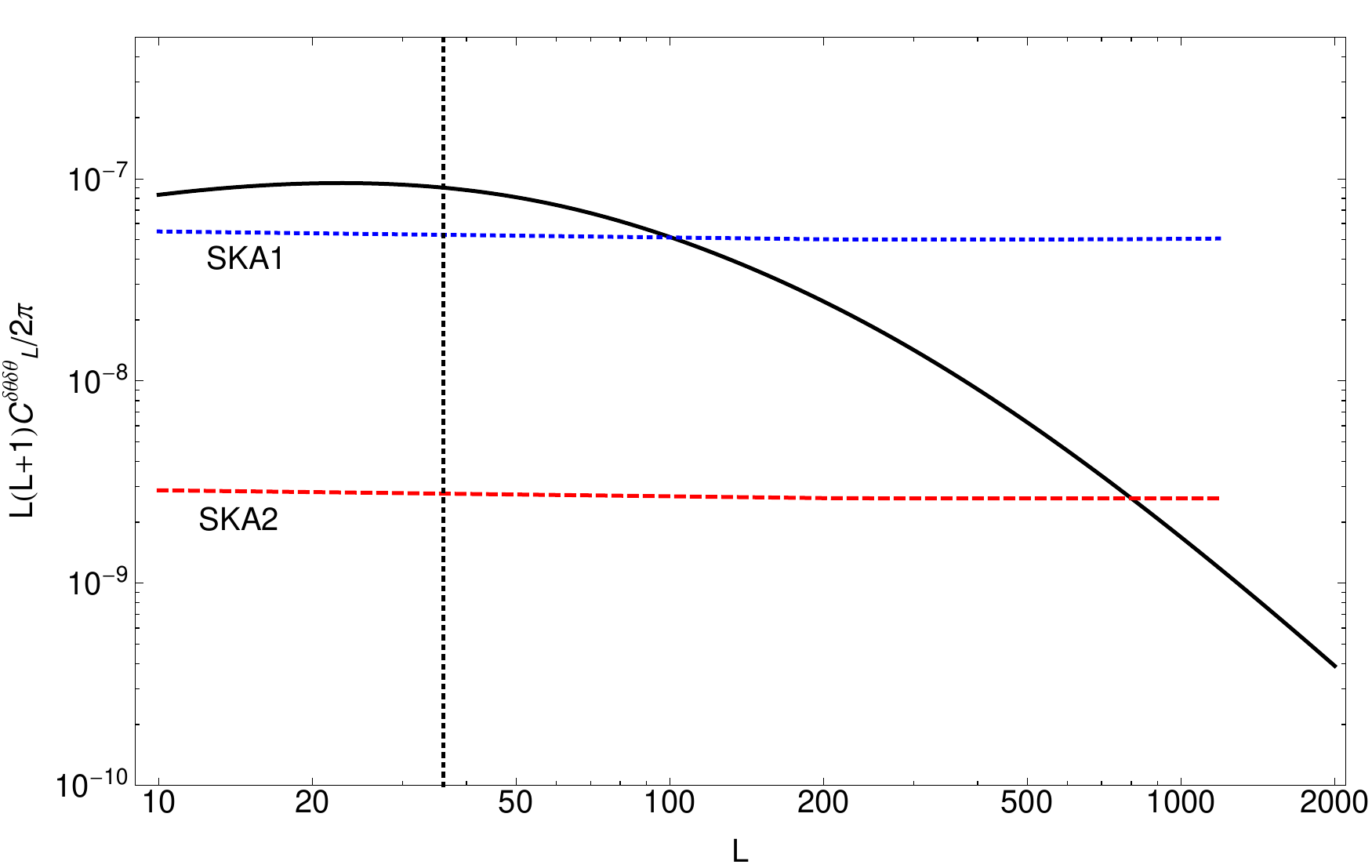}
}
\caption{The lensing displacement field power spectrum, $C^{\delta\theta \delta\theta}_L$, for sources at $z=8$ is shown as a solid black line and lensing reconstruction noise $N_L$ as dashed/dotted lines.  The blue dotted curve is for the SKA1 Baseline Design with 10 8~MHz frequency bins around $z=8$ spanning the redshift range $z \simeq 6.5-11$.  The red dashed line is for SKA2 and the same frequency bins. The vertical line is approximately the lowest $L$ accessible with a 5-by-5 degree field.  Where the noise curves are below $C_L$, typical fluctuations in the lensing deflection should be recoverable in a map. }
\label{fig:CLNL}
\end{figure}

At redshift $z_s \sim 8$, we assume the SKA1 Baseline Design \citep{Dewdney:2013} parameters of $A_{\rm coll} \simeq 0.3  \, {\rm km}^2$ with maximum baseline $D_{\rm tel}=4 \, {\rm km}$, while for SKA2 we can consider $A_{\rm coll} \simeq 1.2  \, {\rm km}^2$.
The estimated lensing noise is shown in Figure~\ref{fig:CLNL} along with the estimated signal.  
Here $C^{\delta \theta \delta \theta}_L$ is the convergence field power spectrum at $z_s=8$ and $N_L$ the lensing reconstruction noise assuming a reionization fraction $f_{\rm HI}=1$. 
Note that at such high redshifts the system temperature $T_{\rm sys}$ is dominated by galactic synchrotron radiation and can be approximated by  \citep{Dewdney:2013}
\begin{equation}
T_{\rm sys} = 60 \times \left(\frac{\nu}{300 \,{\rm MHz}}\right)^{-2.55} \, {\rm K}.
\end{equation}

For SKA\_Low we can consider a $1,000~{\rm hr}$ observation time and multiple stacked bands $\nu$ with $B = 8 \, {\rm MHz}$\footnote{As discussed in \citep{Metcalf:2009}, the quadratic lensing estimator is optimized for the case where the statistical properties of the 21cm radiation signal and noise are constant within a band, and an observation bandwidth of a few ${\rm MHz}$ is small enough so that this assumption is justified.}. Here we note that the noise $N(L,\nu)$ converges quickly with $j$ (as going to higher values of $j$ the signal decays quickly below the thermal noise level), so that $j_{\rm max} \sim 20$ at $z \sim 8$.

It can be seen in Fig.~\ref{fig:CLNL} that the noise for at least SKA2 should be well below the power-spectrum over a large range in angular scales.  This means that typical modes will be measurable with high signal-to-noise and a high fidelity map is possible.  The noise crosses the power-spectrum at $l\sim 800$ for SKA2 which corresponds to an angular resolution of $\sim 10$ ~ arcmin. This measurement greatly benefits from the larger collecting area that will come with Phase 2 of the SKA.  Such a high fidelity map could be cross-correlated with many other observables from lower redshift to determine how gas and stars follow dark matter as a function of redshift or to test the predictions of general relativity.

\section{Weak Lensing with HI Intensity Mapping after Reionization}
\label{sec:LensIM}

After reionization HI gas exists only in discrete galaxies.  This introduces some differences in the calculation of the lensing estimator.  As will be seen, the noise per mode will increase from the EoR case, but because planned 21~cm intensity mapping surveys will cover a much larger fraction of the sky the sample variance is much lower and a very good measurement of the convergence power-spectrum and its redshift evolution should be possible.  The signal-to-noise is critically dependent on the array parameters, the area surveyed and the evolution of the HI mass fraction with redshift. 

\subsection{Measuring lensing with the SKA}

In previous work \citep{PourtsidouMetcalf:2014} we extended the 21~cm intensity mapping lensing method further, taking into account the discreteness and clustering of galaxies, and investigated the possibility of measuring lensing at intermediate redshifts without resolving (in angular resolution) or even identifying individual sources. Here we perform an improved analysis of the signal-to-noise expected from an SKA-like interferometer, and demonstrate the performance of SKA\_Mid Phases 1 and 2.   A formal treatment of how a quadratic lensing estimator can be derived for discrete sources is included in appendices~\ref{sec:intens-mapp-lens} and \ref{sec:lens-estim-clust}.

The expected measurement error
in the power spectrum $C_L^{\delta \theta \delta \theta}$ in a band of width $\Delta L$ is given by 
\begin{equation}
\label{eq:DCL}
\Delta C_L^{\delta \theta \delta \theta} =\sqrt{\frac{2}{(2L+1)\Delta L f_{\rm{sky}}}}\left(C_L^{\delta \theta \delta \theta}+N_L\right).
\end{equation} 
Here, $N_L$ is the derived $\delta \theta$ estimate reconstruction noise (see  Appendix~\ref{sec:lens-estim-clust} for the detailed derivation and formulas for this case) which involves the underlying dark matter power spectrum, the HI density $\Omega_{\rm HI}(z)$
as well as the HI mass moments up to 4th order and, of course, the thermal noise of the array $C^{\rm N}_\ell$, equation \eqref{eq:CellN}. 
The system temperature in this case is the sum of the receiver and sky noise \citep{Dewdney:2013}
\begin{equation}
T_{\rm sys} = \left[40+1.1\times 60 \times \left(\frac{\nu}{300 \,{\rm MHz}}\right)^{-2.55}\right] \, {\rm K}.
\end{equation} 

The moments of the HI mass function (or 21~cm luminosity function) enter into the estimator's noise expressions \eqref{eq:I0cont} through \eqref{eq:I4cont}.  Assuming that
the HI mass function follows the Schechter function (which is an excellent fit to the local Universe data --- see, for example, the HIPASS survey results in \citep{Zwaan03}) we can write
\begin{equation}
\frac{dn(M,z)}{dM}dM = \phis(z) \left(\frac{M}{M^\star(z)}\right)^\alpha {\rm exp}\left[-\frac{M}{M^\star(z)}\right]\frac{dM}{M^\star(z)}.
\end{equation} 
Damped Lyman-$\alpha$ systems observations are used to measure $\Omega_{\rm HI}(z)$ up to $z \sim 5$, while no measurements of $\phis(z)$ and $M^\star(z)$ are available other than in the local Universe. However, for this form of the mass function we have the relation
\begin{equation}
\label{eq:omHI}
\Omega_{\rm HI}(z) = 4.9 \times 10^{-4} \frac{\phis(z)}{\phis(z=0)}\frac{M^\star(z)}{M^\star(z=0)}.
\end{equation}
At redshift $z=0$ we use the values
$\alpha=-1.3, M^\star=3.47h^{-2}10^9 \, M_\odot, \phis = 0.0204 \, h^3
\, {\rm Mpc}^{-3}$ reported from the HIPASS survey \citep{Zwaan03}. 
We will initially consider the most conservative scenario, i.e. a no evolution model that we will call  Model A.
Model A was also used in \citep{PourtsidouMetcalf:2014}. Here we will also show results using a few HI evolution models.

Let us first concentrate on source redshift $z_s \sim 2$. 
In the recently published SKA1 Baseline Design \citep{Dewdney:2013}, SKA\_Mid operates in the frequency range $\sim 350-3050 \, {\rm MHz}$ divided in three bands. Our chosen $z_s$ corresponds to frequency $\nu_s=473 \, {\rm MHz}$ (Band 1).
Following the proposed parameters for SKA\_Mid we consider a $2 \, {\rm yr}$ observation time, $f_{\rm sky}\sim 0.7$, and we further choose $B = 20 \, {\rm MHz}$ and $\Delta L=36$. 
From Eq.~(\ref{eq:Saver}) we see that the measurement errors increase as $1/\sqrt{f_{\rm sky}}$ and the fact that  SKA\_Mid will cover a large fraction of the sky contributes to the high S/N significance we predict. An increase in the observation time would reduce the thermal noise of the interferometer and would also lead to an increase of the overall S/N levels. A similar effect comes from increasing the bandwidth $B$ of the observation --- however, note that we avoid choosing very wide bands since in that case there would be non-negligible correlations within the band. It is not clear whether these correlations would just lead to an increase of the noise only or if they would also increase the lensing signal. We plan to investigate this issue in a future publication.

\begin{figure}
\centering
\includegraphics[scale=0.55]{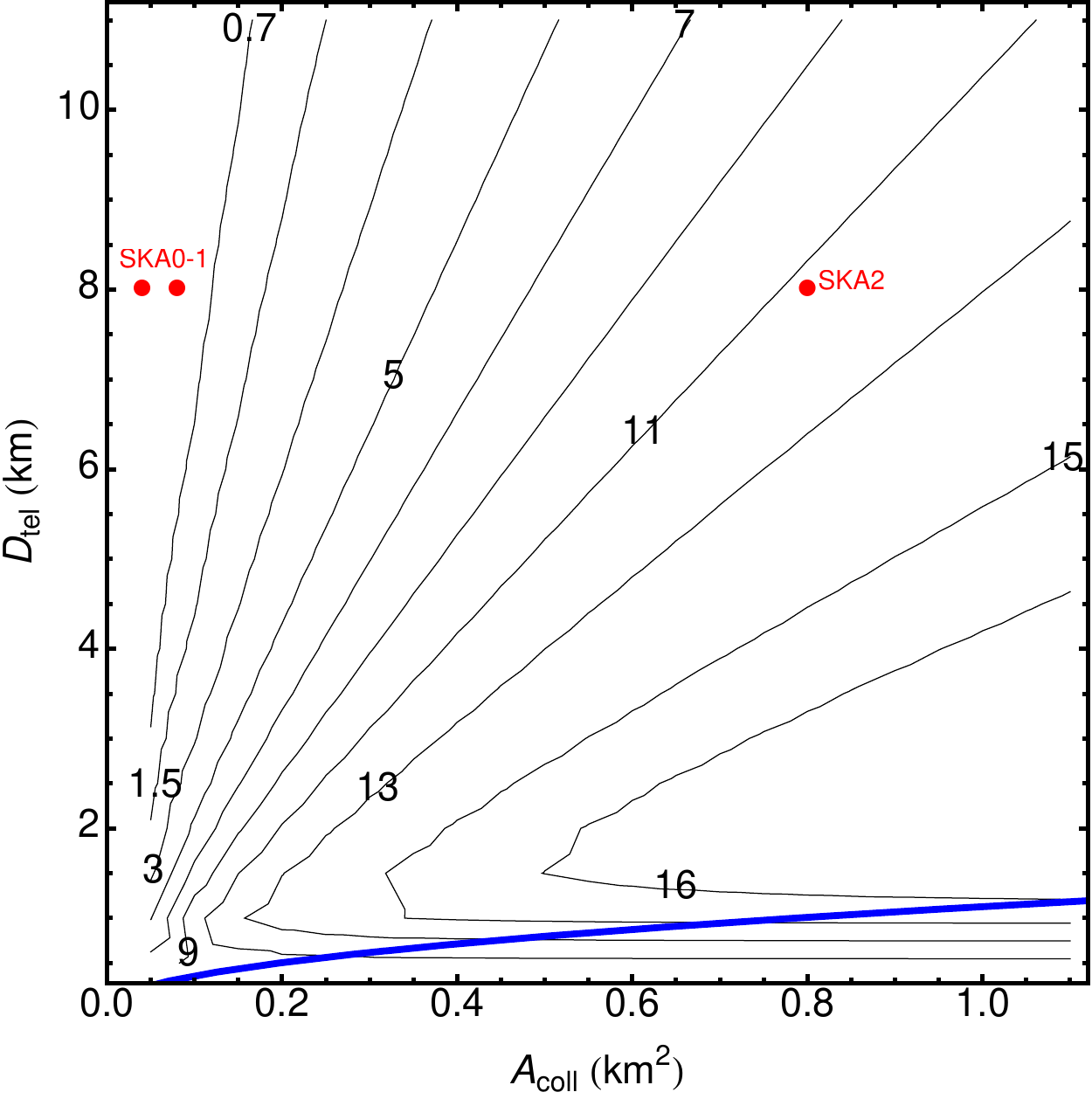}
\caption{The signal-to-noise at $L=100$ in a $\Delta L = 36$ for various
 array configurations. Sources are at $z_s=2$. The contour lines are labelled with the
  (S/N) values. The area under the thick blue solid line is excluded, since
  it corresponds to $f_{\rm cover} > 1$. 
  SKA0, SKA1 and SKA2 are shown. The no-evolution in $\Omega_{\rm HI}$ (Model A) scenario is used.}
\label{fig:SoverN}
\end{figure}

\begin{figure}
\centering
\includegraphics[scale=0.35]{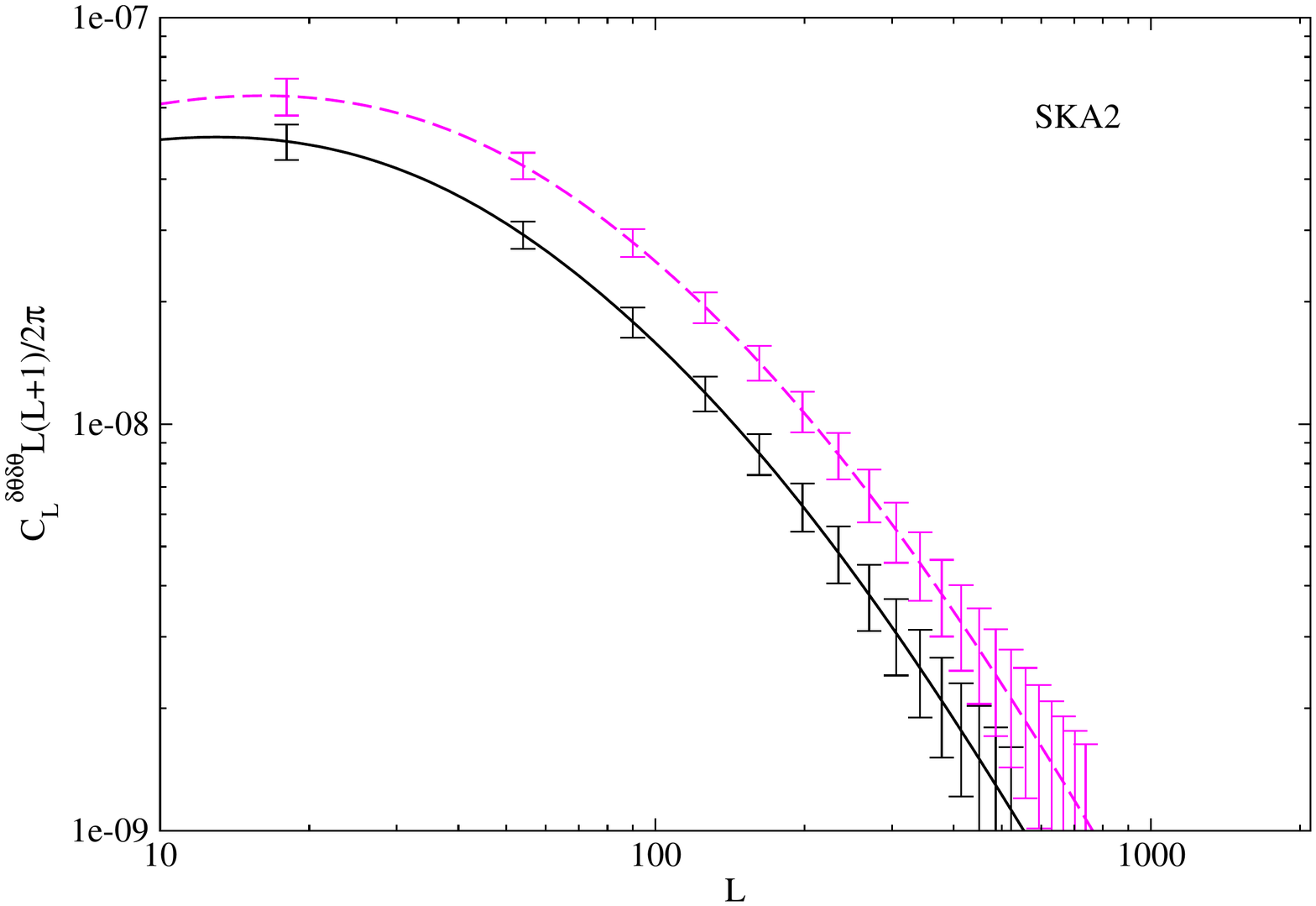}
\caption{Displacement field power spectrum for $z_s=2$ (solid black line) and $z_s=3$ (dashed magenta line) and the corresponding measurement errors using the SKA2 specifications and Model A for the HI mass function. }
\label{fig:CL}
\end{figure}

Since the galaxies are being treated as point sources we wish to include only frequency modes that do not resolve the internal structure of the galaxies.
 Assuming a typical velocity dispersion for a galaxy at $z=2$ to be around $200 \, {\rm km/sec}$, we find $j_{\rm max}=63$. However the noise has converged at $j \sim 40$, which is the value we use in our numerical simulations (see Fig. \ref{fig:CLmodelBjconv} for a demonstration of the convergence with $j$). Keeping these values constant, we present a signal-to-noise contour plot (Figure \ref{fig:SoverN}) at multipole $L=100$ for $z_s=2$ on the $(A_{\rm coll},D_{\rm tell})$ parameter space, and show three SKA\_Mid performance cases: SKA\_Mid Phase 1 with $50\%$ sensitivity (SKA0), SKA\_Mid Phase 1 (SKA1), and SKA\_Mid Phase 2 (SKA2). We also present the displacement field power spectrum and measurement errors for $z_s=2$ and $z_s=3$ using the SKA2 specifications in Figure \ref{fig:CL}. These results are not sensitive to the exact value of  $\ell_{\rm min}$. For example, the S/N values remain practically unchanged if we use a very small $\ell_{\rm min}=10$ or the one corresponding to the current SKA1 field of view ($\sim 1 \, {\rm deg}$), $\ell_{\rm min}=180$.

\begin{figure}
\centering
\includegraphics[scale=0.35]{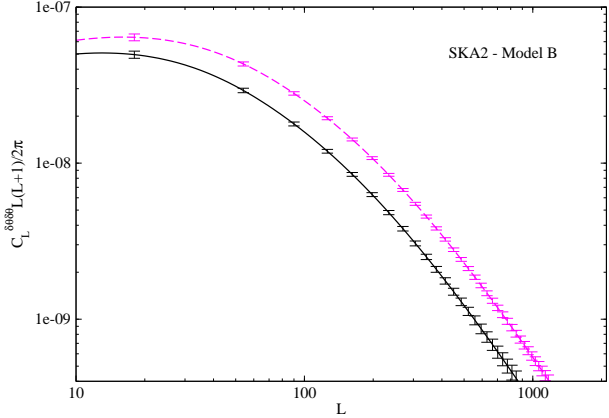}
\caption{Displacement field power spectrum for $z_s=2$ (solid black line) and $z_s=3$ (dashed magenta line) and the corresponding measurement errors using the SKA2 specifications and Model B for the HI mass function.}
\label{fig:CLmodelB}
\end{figure}

This technique should enable us to measure the lensing power spectrum at source redshifts well beyond those accessible with more traditional weak lensing surveys based on the shearing of individual galaxy images.  The noise is expected to be small enough that evolution in the lensing power spectrum between $z_s=2$ and $3$ will be clearly detectable as illustrated in Figure~\ref{fig:CL}.

\subsection{Importance of the redshift evolution in the HI mass function}

It is useful to illustrate how our results depend on a possible evolution of the HI mass function with redshift.
We will follow \cite{Zhang06}. The authors construct three different evolution models for the HI mass function. Observations of damped Lyman-$\alpha$ systems and Lyman-$\alpha$ limit systems measure $\Omega_{\rm HI}$ from $z=0$ to $z \sim 5$ (see, for example, \citep{Peroux:2001ca}).These observations found that
$\Omega_{\rm HI}$ increases by a factor of $5$ toward $z\sim 3$ and then decreases toward higher redshift.   Combining Eq.~(\ref{eq:omHI}) and these observations, one can put constraints on the product of $\phis$ and $M^\star$.  Either an increase in $\phis$ or $M^\star$ increases the detectability of lensing, so the no evolution model is considered a conservative choice.

\begin{figure}
\centerline{
\includegraphics[scale=0.5]{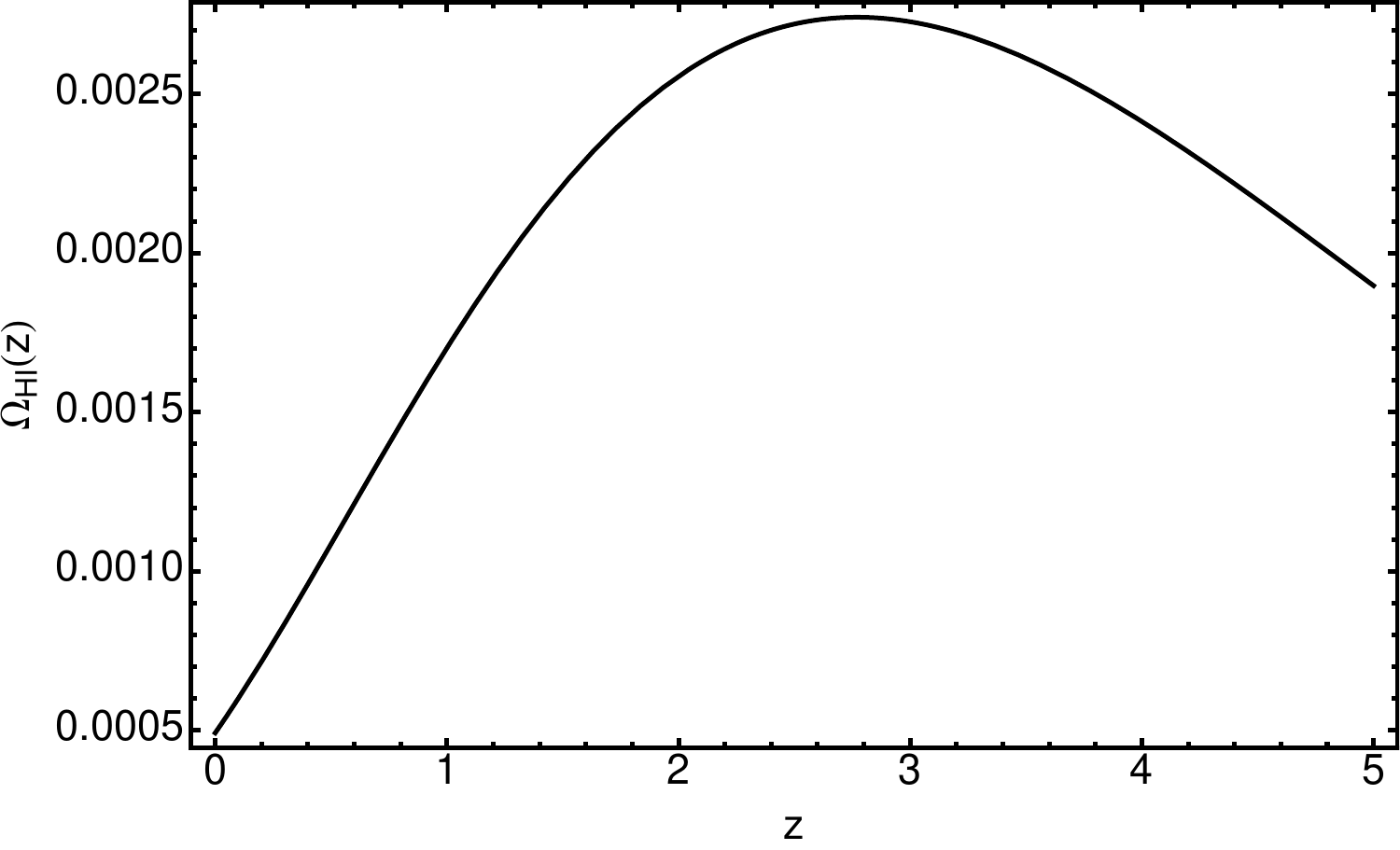}}
\caption{The evolution of $\Omega_{\rm HI}(z)$ with redshift in \citep{Zhang06}.}
\label{omHImodelA}
\end{figure}

Following \citep{Zhang06}, the evolution of $\Omega_{\rm HI}$ will be approximated as 
\begin{equation}
g(z) \equiv \frac{\Omega_{\rm HI}(z)}{\Omega_{\rm HI}(z=0)} = (1+z)^{2.9} \, {\rm exp}(-z/1.3).
\end{equation} This function is plotted in Fig.~\ref{omHImodelA} and the constraint for the redshift evolution of the product $\phis M^\star$ is
\begin{equation}
\phis(z)M^\star(z)=\phis(z=0)M^\star(z=0)g(z).
\end{equation}
Here we will investigate the following evolution scenarios \citep{Zhang06}: 
\begin{itemize}
\item {\bf Model (B)} No evolution in $M^\star(z)$. $\phis(z)=\phis(z=0)g(z)$. 
\item {\bf Model (C)} No evolution in $\phis(z)$. $M^\star(z)=M^\star(z=0)g(z)$. 
\item {\bf Model (D)} $\phis(z)/\phis(z=0)=M^\star(z)/M^\star(z=0)=g(z)^{1/2}$.
\end{itemize}
We also note that HI mass function modeling has been 
performed by other authors \citep{Abdalla:2009wr}, where the evolution of the $\Omega_{\rm HI}(z)$ function agrees with  \citep{Zhang06} until $z \sim 3$ and so would make very little difference to our predictions.  

\subsubsection{Model B}

\begin{figure}
\centering
\includegraphics[scale=0.55]{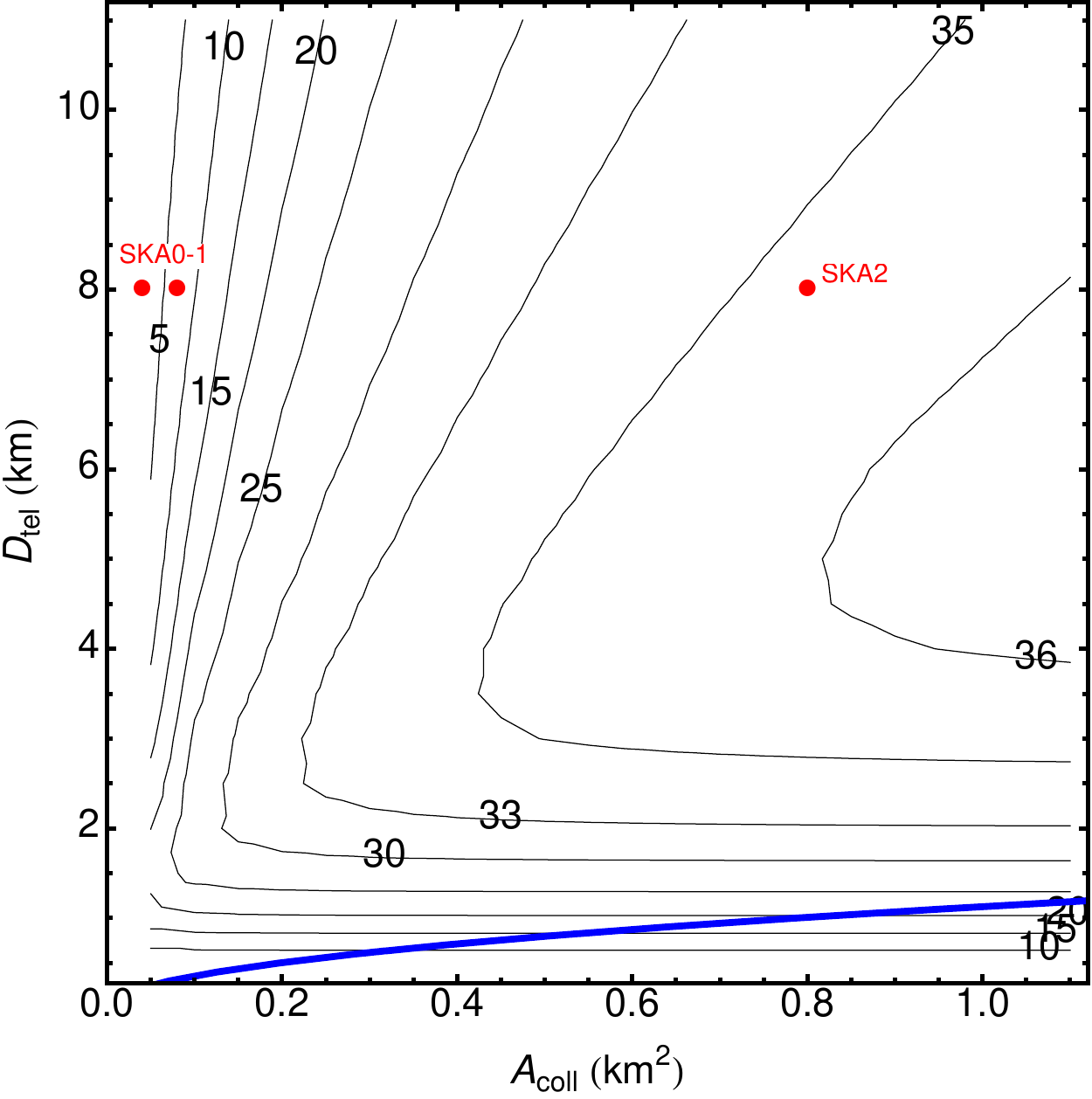}
\caption{The signal-to-noise at $L=100$ in a $\Delta L = 36$ for various
 array configurations.  Model B for the evolution of the HI mass function is used. Sources are at $z_s=2$. The contour lines are labelled with the
  (S/N) values. The area under the thick blue solid line is excluded, since
  it corresponds to $f_{\rm cover} > 1$. 
  SKA0, SKA1 and SKA2 are shown. }
\label{fig:SoverNmodelA}
\end{figure}

Model B assumes no evolution in $M^\star$.   The shot noise terms in the estimator noise are changed in this case.  The displacement field power spectrum and measurement errors for model B are shown in Fig.~\ref{fig:CLmodelB}, and the signal-to-noise contour at $L=100$ is shown in Fig.~\ref{fig:SoverNmodelA} as a function of array parameters. Notice the improvement of the S/N across all of the parameter space from the no-evolution model--- the lensing signal can be detected even with SKA0.

\subsubsection{Model C}

Model C assumes no evolution in $\phi^\star$. 
$\Omega_{\rm HI}$ will increase, but the shot noise terms will stay the same as the no-evolution ones (Model A), because the integrals do not get affected by $M^\star$. Hence, the signal-to-noise is higher than Model A by a factor $\sim 1.5$ at $L=100$ using the SKA2 specifications, but lower than model B by a factor $\sim 0.5$ at $L=100$ using the SKA2 specifications.

\subsubsection{Model D}

Model D assumes that both $\phis$ and $M^\star$ are evolving as $\phis(z)/\phis(z=0)=M^\star(z)/M^\star(z=0)=g(z)^{1/2}$. The signal-to-noise is higher than model C by a factor $\sim 1.5$ at $L=100$ using the SKA2 specifications, but lower than model B by a factor $\sim 0.7$ at $L=100$ using the SKA2 specifications.

We see that for all models where the HI density increases with redshift the lensing signal-to-noise is greater than in the no-evolution model.  Furthermore,  
we learned that there is a stronger dependence on the evolution of $\phis$ than $M^*$. This is also expected, since an increase in $\phis$ increases the detectability of lensing --- at this point we should stress that in our case the contributions of the Poisson moments contribute both to the signal as well as the noise of our estimator, so the dependence on $\phis$ is crucial. Model B, in which $\phis$ is solely responsible for the increase in $\Omega_{\rm HI}$ is the most optimistic scenario, but the most conservative no-evolution Model A is probably less realistic.

\begin{figure}
\centerline{
\includegraphics[scale=0.55]{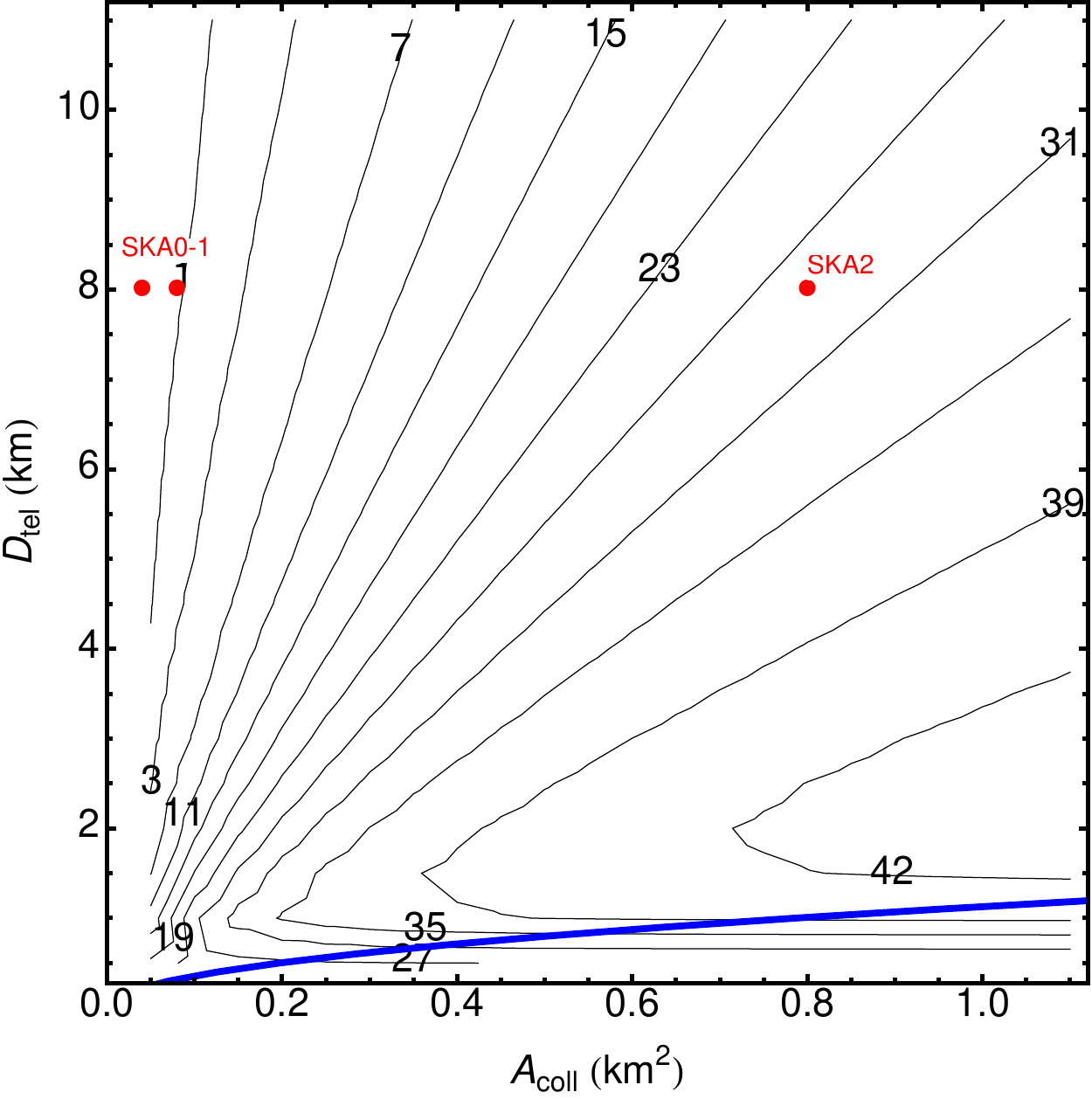}}
\caption{The signal-to-noise from Eq.~(\ref{eq:SovNband}) for various
  array configurations. Sources are at $z_s=2$ and we use Model A for the HI mass function.  The contour lines are labelled with the
  $(S/N)$ values. The area under the thick blue solid line is excluded, since
  it corresponds to $f_{\rm cover} > 1$.  SKA0, SKA1 and SKA2 are shown.}
\label{contour2}
\end{figure}

\begin{figure}
\centerline{
\includegraphics[scale=0.55]{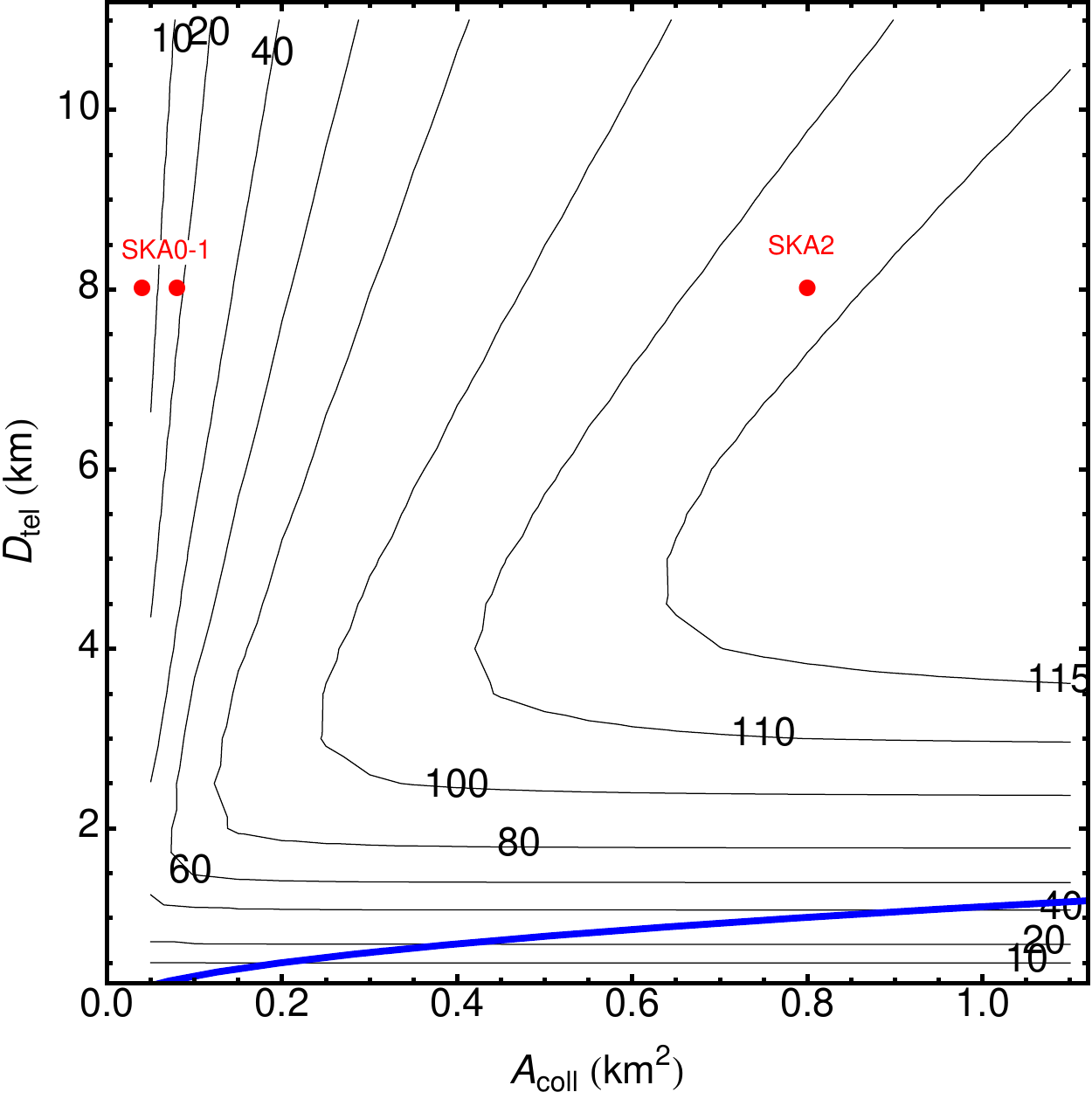}}
\caption{The signal-to-noise from Eq.~(\ref{eq:SovNband}) for various
  array configurations. Sources are at $z_s=2$ and we use Model B for the HI mass function. The contour lines are labelled with the
  $(S/N)$ values. The area under the thick blue solid line is excluded, since
  it corresponds to $f_{\rm cover} > 1$.  SKA0, SKA1 and SKA2 are shown.}
\label{contour2modB}
\end{figure}

Previously, we used the signal-to-noise calculations at a single multipole number ($L=100$) to demonstrate the 
capabilities of various arrays (see Fig.~\ref{fig:SoverN} and \ref{fig:SoverNmodelA}). A more comprehensive expression would involve contributions from many multipoles. We therefore define a new (S/N) for the detection of the lensing signal as
\begin{equation}
\label{eq:SovNband}
{\rm S/N} = \left(\sum_{L_b} \frac{C^2_{L_b}}{\Delta C^2_{L_b}}\right)^{1/2},
\end{equation} 
where the sum is over deflection field power spectrum band powers (using $\Delta L=36$ as before) and $\Delta C_{L_b}$ is the forecasted error on each band power. We calculate (S/N) as a function of collecting area and maximum baseline, with $L_{b_{\rm min}}=18$ and $L_{b_{\rm max}}=990$. We show the signal-to-noise contour plots for Model A and model B for the HI mass functions in Figures~\ref{contour2} and \ref{contour2modB}, respectively. 

\begin{figure}
\centering
\includegraphics[scale=0.35]{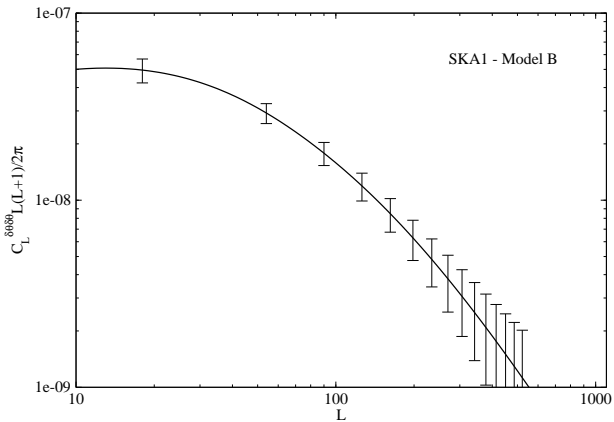}
\caption{ Displacement field power spectrum for $z_s=2$ and the corresponding measurement errors using the SKA1 specifications and Model B for the HI mass function.}
\label{fig:CLmodelBska1}
\end{figure}

As we have already stressed, Model A is the most conservative, no-evolution scenario one can consider for modeling the HI mass function. From Figures~\ref{fig:CL} and \ref{contour2} we can deduce that in order to get a good measurement of the lensing signal up to multipole number $L \sim 1000$ we need a S/N of about $25$. Then from Fig.~\ref{contour2modB} we can see that SKA1 assuming Model B does almost as well as SKA2 assuming no evolution (sources at $z_s=2$). This shows how crucial the HI mass function evolution is for our method and the SKA's science goals. In Fig.~\ref{fig:CLmodelBska1} we show the results of our calculations using the SKA1 specifications and Model B --- as expected from the above discussion, the measurement errors are similar to the SKA2 no-evolution case.

\section{Constraining Interacting Dark Energy Models}
\label{Sec:coupledDE}

In this Section we demonstrate the constraining power of 21cm lensing measurements using the intensity mapping method with an instrument like the SKA. 
The problems of the nature and evolution of dark energy and dark matter are arguably the most important ones in modern cosmology, and ambitious future missions like EUCLID\footnote{http://sci.esa.int/euclid/} are dedicated to the exploration and mapping of the dark Universe by investigating the evolution of cosmic structures out to redshifts $\sim 1.5-2$, a period over which dark energy is thought to dominate the Universe's expansion.

An important question is whether dark energy is indeed a cosmological constant, like all the available data from Cosmic Microwave Background measurements \citep{Ade:2013}, the Hubble constant \citep{Riess} luminosity and distance at high redshift with supernovae Ia \citep{Kowalski:2008ez}  and Baryon Acoustic Oscillation surveys \citep{Lampeitl:2009jq} seem to suggest (however this success comes with the price of the cosmological constant and coincidence problems), an evolving scalar field like quintessence, or a modification of gravity. To differentiate between the plethora of available models, we might have to go even deeper in redshift space $z > 1.5$, and this is where 21~cm radiation becomes a unique probe of cosmology.

Interactions in the dark sector, such as a non-gravitational coupling between dark matter and dark energy (for example a coupling between a quintessence field $\phi$ playing the role of dark energy and the matter sector, see \citep{Amendola2000}), can cause modifications to the background evolution of the matter density and the Hubble parameter, as well as changes in the evolution of structure growth with respect to $\Lambda$CDM. However, as we will demonstrate below, the background effects are very difficult to probe as alternative dark energy and modified gravity models can be very successful in mimicking the $\Lambda$CDM background. Therefore, we need to study linear perturbations and quantify the effects of the various models in the CMB and matter power spectra. Modifications in the evolution of structure growth in interacting dark energy models leave distinctive signatures in the weak lensing signal, and measurements such as those presented in Fig.~\ref{fig:CL}, could be used to constrain such models.

In a recent study \cite{Pourtsidou:2013} constructed three general classes (Types) of models of dark energy in the form of a scalar field $\phi$ coupled to cold dark matter. The first class, Type 1, is a generalization of the coupled quintessence (CQ) model suggested by \citep{Amendola2000}. In such models, the CDM energy density is not conserved separately (like in the uncoupled case) but there is a non-zero coupling current which represents the decay of DM to DE or vice versa. In the specific \citep{Amendola2000} model the evolution of the background CDM density $\rhob_c$ is found to be
\begin{equation}
\rhob_c = \rhob_{c,0}a^{-3}e^{\alpha \phi}.
\end{equation} We therefore see that the CDM density evolves differently than the uncoupled quintessence case and $\Lambda$CDM, which means that for positive coupling parameter $\alpha$ (decay of dark matter to dark energy) the CDM density has to be higher in the past in order to reach the same value today.

This model and its variants have been extensively studied in the literature, and constraints on the strength of the coupling parameter (roughly $< 0.1$ assuming it is constant) have been derived using its effects on the CMB and matter power spectra, the growth of structure and the weak lensing signal (e.g. \citep{Tarrant:2011qe,Xia:2013nua,CalderaCabral:2009ja,DeBernardis:2011iw,Amendola:2011ie} and references therein). 

Here we will investigate the weak lensing convergence signatures of models belonging to the new classes of coupled DE theories, namely Type 2 and Type 3, constructed in \cite{Pourtsidou:2013}. 
Type 2 models involve both energy and momentum transfer between the two components of the dark sector (i.e. like Type 1 models but with a distinctively different coupling mechanism), while Type 3 is a pure momentum transfer theory. The displacement field power spectrum for these models is calculated using \citep{Kaiser:1992}
\bea \nonumber
\label{eq:CLgen}
C_L^{\delta \theta \delta \theta} = \frac{9}{L(L+1) c^3} \int_0^{z_s}dz \, [\Omega_m(z)]^2 P(k=L/{\cal D},z) [W(z)]^2  \\
\times (1+z)^{-4}[H(z)]^3 \, ,
\eea which simplifies to Eq.~(\ref{eq:CL}) for $\Lambda$CDM by setting $\frac{\Omega_m(z)}{\Omega_m}=\frac{H^2_0}{a^3[H(z)]^2}$.

\subsection{Type 2 models of CDM coupled to DE}

Type 2 models are classified via the Lagrangian function \citep{Pourtsidou:2013}
\begin{equation}
L(n,Y,Z,\phi) = F(Y,\phi) + f(n,Z),
\end{equation} where $n$ is the particle number density, $Y=\frac{1}{2}\nabla_\mu \phi \nabla^\mu \phi$ is used to construct a kinetic term for the quintessence field $\phi$, and $Z=u^\mu \nabla_\mu \phi$ plays the role of a direct coupling of the fluid velocity to the gradient of the scalar field. 

We can proceed by making further assumptions. First of all, we want to consider coupled quintessence models, hence we can write $F=Y+V(\phi)$. Secondly, since we choose  the scalar field to be coupled to CDM, the function $f$ is separable, i.e.
$
f=n \, h(Z).
$

The details of the field and fluid equations for the Type-2 class of theories under consideration can be found in \citep{Pourtsidou:2013}. An important quantity that appears in the cosmological equations is the function
$
K(Z)=h_z/(h-Zh_z),
$ with $h_z=dh/dZ$. We therefore see that we cannot simply choose $h$ to be proportional to $Z$, i.e. $h(Z) = \beta Z$ with $\beta$ the (constant) coupling parameter, as in that case $K(Z)$ diverges. 

\begin{figure}
\centering
\includegraphics[scale=0.35]{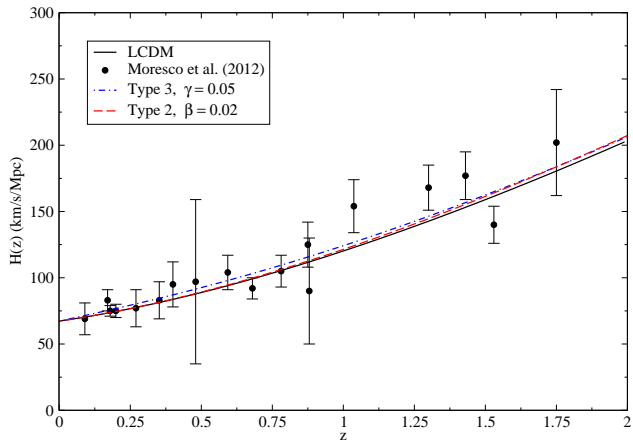}
\caption{The Hubble parameter versus redshift for $\Lambda$CDM (solid black line) and the Type 2, 3 models under consideration (red dashed line for Type 2, blue dotted dashed line for Type 3), together with expansion history measurements from \citep{Moresco:2012by}, which combine the observational constraints on the Hubble parameter from \citep{Simonetal,Stern:2009ep,Moresco:2012jh}.}
\label{fig:HzT2}
\end{figure}

\begin{figure}
\centering
\includegraphics[scale=0.35]{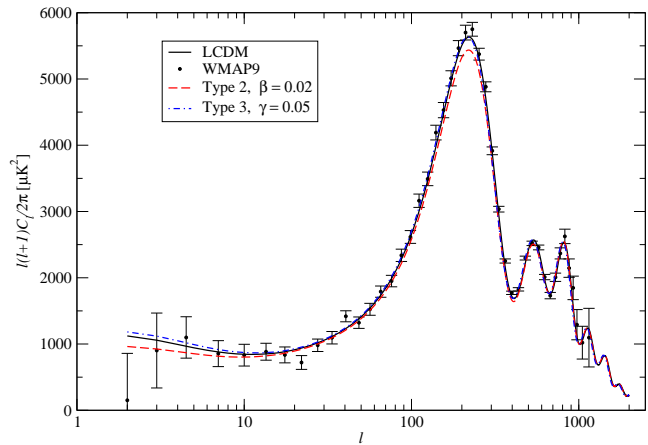}
\caption{The CMB TT spectra for $\Lambda$CDM (solid black line) and the Type 2, 3 models under consideration (red dashed line for Type 2, blue dotted dashed line for Type 3), together with WMAP9 measurements.}
\label{fig:CMBT2}
\end{figure}

Choosing $h(Z) = Z^\beta$ we find that the background CDM density $\bar{\rho}_c$ evolves as
\begin{equation}
\bar{\rho}_c=\bar{\rho}_{c,0}a^{-3}\bar{Z}^\beta,
\end{equation} where $\bar{Z}=-\dot{\bar{\phi}}/a$ (note that the dot denotes derivatives with respect to conformal time). The fact that the CDM density depends on the time derivative $\dot{\bar{\phi}}$ instead of $\bar{\phi}$ itself is a notable difference from the Type 1 (Amendola) case. Note also that the specific form of the coupling function $Z^\beta$ 
means that in order to have meaningful solutions $\bar{Z}$ has to be positive throughout the cosmological evolution. 
In \citep{Pourtsidou:2013} the authors derive the background and perturbed Klein-Gordon equations for the evolution of the quintessence field, as well as the perturbed CDM equations for the evolution of the density contrasts and velocities.
With these equations at hand, and considering a single exponential potential for the quintessence field, we use a modified version of the \texttt{CAMB} code \citep{camb} to study the observational signatures of the chosen coupled model. In Figures~\ref{fig:HzT2} and \ref{fig:CMBT2} we show the Hubble parameter evolution and the CMB TT power spectra for these models together with $\Lambda$CDM. We also construct the displacement field power spectrum (Fig.~\ref{fig:CLcoupled}) and compare it with the $\Lambda$CDM prediction in Fig.~\ref{fig:CL}. Note that the cosmology of each model evolves to the PLANCK cosmological parameter values at $z=0$ (Ade et al. 2013) and the matter power spectra $P(k)$ have been normalized to CMB fluctuations. 

\begin{figure}
\centering
\includegraphics[scale=0.55]{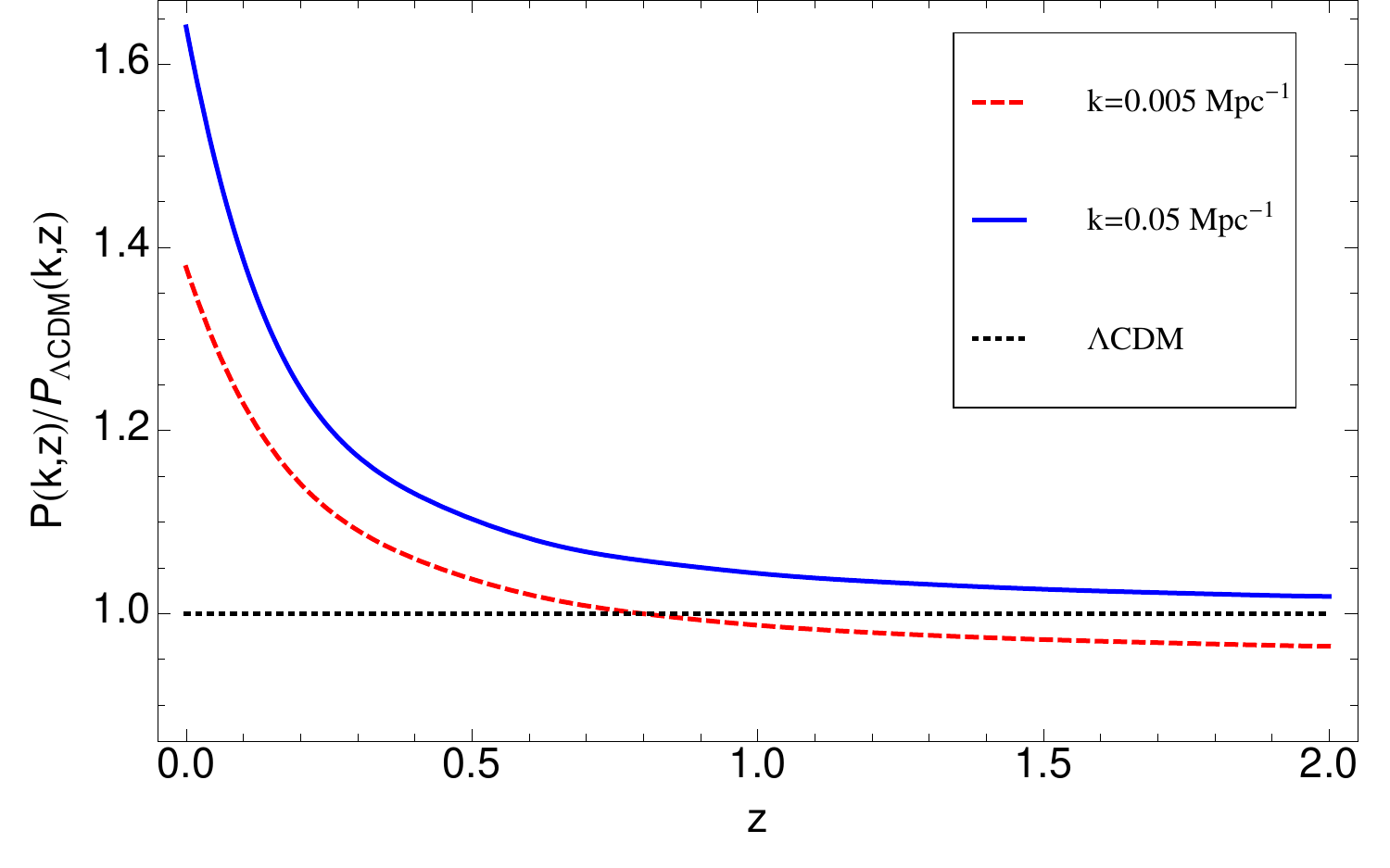}
\caption{The ratio of the linear matter power spectrum in the Type 2 coupled DE model to the one in $\Lambda$CDM for two characteristic scales as a function of redshift.}
\label{fig:growthT2}
\end{figure}

From Fig.~\ref{fig:CLcoupled} we see that the Type 2 model with a coupling parameter $\beta=0.02$ would be excluded by the proposed lensing / intensity mapping observations. That is because in this case there is energy transfer from dark matter to dark energy making the dark matter density larger in the past compared to the non-interacting case for fixed $\Omega_m$ today, the growth is increased, the gravitational potential is higher and the convergence power spectrum is enhanced with respect to $\Lambda$CDM. 
Another important point demonstrated in Fig.~\ref{fig:growthT2} is that the Type 2 model linear growth factor is strongly scale dependent, and the difference between the growth of the two scales persists up to the source redshift $z=2$. Note that there is also an effect on the Hubble parameter evolution (see Fig.~\ref{fig:HzT2}) and the CMB power spectrum (see Fig.~\ref{fig:CMBT2}), but subtler than the effect on the lensing signal.

\subsection{Type 3 models of CDM coupled to DE}

Type 3 models are classified via the Lagrangian function \citep{Pourtsidou:2013}
\begin{equation}
L(n,Y,Z,\phi) = F(Y,Z,\phi) + f(n).
\end{equation}
We can consider a coupled quintessence function of the form
$F=Y+V(\phi)+h(Z)$, and choose $h(Z)=\gamma Z^2$. Note that this model is physically acceptable 
for $\gamma < 1/2$, otherwise it suffers from ghost and strong coupling problems \citep{Pourtsidou:2013}.

As we have already mentioned, Type 3 is a pure momentum transfer theory. The background CDM density $\rhob_c$ evolves as in the uncoupled case and the energy conservation equation remains uncoupled even at the linear level.
Following the same procedure, using the equations derived in \citep{Pourtsidou:2013} and the modified version of \texttt{CAMB} we construct the displacement field power spectrum and compare it with the $\Lambda$CDM prediction in Fig.~\ref{fig:CL}. 

\begin{figure}
\centering
\includegraphics[scale=0.35]{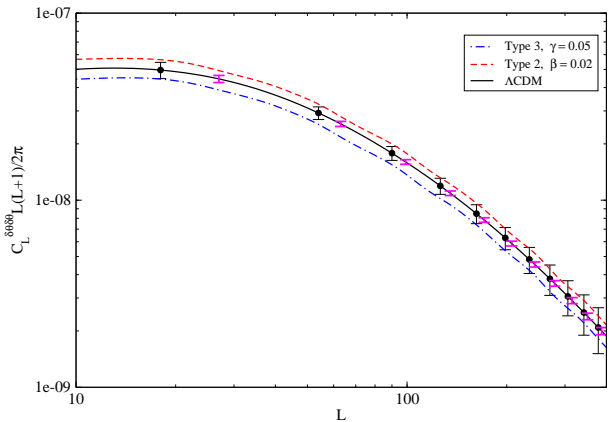}
\caption{Displacement field power spectrum for $\Lambda$CDM (solid black line) compared with Type 2 (red dashed line) and Type 3 (blue dotted dashed line) interacting dark energy models. Sources are at $z_s=2$. The circle points with black error bars are the same as in Figure~\ref{fig:CL},  i.e. using the most conservative scenario (no evolution in the HI mass function), Model A. Assuming a more optimistic scenario like Model B (magenta error bars) would further improve the constraints. } 
\label{fig:CLcoupled}
\end{figure}

From Fig.~\ref{fig:CLcoupled} we see that the Type 3 model with a coupling parameter $\gamma=0.05$ would be excluded. The signature of this model is growth suppression and hence a decrease in the lensing signal. Again, we note that there is also an effect on the Hubble parameter evolution (see Fig.~\ref{fig:HzT2}) and the CMB power spectrum (see Fig.~\ref{fig:CMBT2}), but subtler than the effect on the lensing signal.

\section{conclusions}
\label{sec:conc}

HI intensity mapping is an innovative technique which can be utilized to map the large scale structure of the Universe in three dimensions. 
In this paper, we investigated the potential of a SKA-like interferometer to perform accurate measurements of the lensing signal over a wide range of redshifts using the intensity mapping method. Extending the work in \citep{PourtsidouMetcalf:2014}, we presented our technique for measuring gravitational lensing in HI observations after reionization. Other methods have been proposed based on counting the number of detected galaxies in 21~cm emission \citep{Zhang05,Zhang06,Zhang11}.  In such methods the clustering of galaxies and the shot noise from Poisson fluctuations in the number counts contribute purely to the noise in the lensing estimator.  The corresponding errors can be controlled to $\sim 10\%-20\%$ level for intermediate source redshifts $z \sim 1$, but for $z_s \gtrsim 2$ the reconstruction is not successful \citep{Zhang11}.
Our estimator takes into account both the clustering and discreetness of galaxies, which contribute to both the noise and to improving the signal resulting in a significant improvement in performance. 

Our calculations showed that the lensing convergence can be measured with high fidelity at redshifts $z\sim 2-3$, hence it can be used for tomographic studies along many redshift (or frequency) bins in order to map, for example, the evolution of the growth function at intermediate redshifts, i.e. higher than those of galaxy shear surveys. We also extended our calculations to include a possible evolution of the HI mass function, showing that the less likely no-evolution scenario is the most conservative one and our claim that a high S/N measurement can be achieved using this method is robust. To quantify the power of this technique we performed an optimization study using a SKA-like interferometer and showed that SKA will be able to deliver very good measurements of the lensing signal and its redshift evolution. We also demonstrated the constraining power of our technique by using it to distinguish specific interacting dark energy models. Our results confirm that the 21cm technique can be used to compete (and complement) with future galaxy surveys, and SKA in particular can be used to investigate the evolution of dark energy and dark matter  in order to test the standard $\Lambda$CDM paradigm against exotic models with possible observational signatures in higher redshifts than the ones probed by  galaxy lensing surveys. 

Another exciting prospect is the study of the reionization epoch --- we showed that the low frequency SKA instrument can map the lensing convergence and allow us to ``see'' the distribution of dark matter in a typical region of the sky, something that is only possible with galaxy lensing around very atypical, large galaxy clusters. This would provide a great opportunity to correlate visible objects with mass and test the dark matter paradigm.
 
\vspace{0.3cm} 
\leftline{\bf Acknowledgments} 
This research is part of the project GLENCO, funded under the Seventh Framework Programme, Ideas, Grant Agreement n. 259349. The authors would like to thank Leon Koopmans, Jonathan Pritchard and Mario Santos for useful discussions.

\bibliographystyle{natbib}

\appendix
\onecolumn

\section{Lensing estimator for a continuous source}
\label{app:intens-mapp-lens-continious}

Here we derive the least squares, or minimum variance, quadratic estimator on a discrete grid for a source that can be approximated as a Gaussian random field.  This will be the discrete version of the estimator found by \citet{Zahn:2005ap}.   The discrete version is necessary for calculations on a finite grid and useful in the deviation of the discrete source case (see Appendix~\ref{sec:intens-mapp-lens}).  This appendix will also serve to introduce some notation.

We write the discrete Fourier transform (DFT)  of the intensity field $I(\bx)$ as
\begin{equation}
I_{\bk} = \frac{\Omega_s}{\Npe \Npa} \sum_{\bx}e^{i\bk \cdot \bx} I(\bx),
\end{equation} where $\bk=(\bell,j)$, $\bx=(\theta,z)$ and $\Omega_s = \Theta_s \times \Theta_s$ for a square
survey geometry. We also have
\begin{equation}
I(\bx) = \frac{1}{\Omega_s}\sum_{\bk}e^{-i\bk \cdot \bx}I_{\bk}.
\end{equation}
$\Npa$ is the number of cells in the direction parallel to the line-of-sight and $\Npe$ is the number of cells on a plane perpendicular to the line-of-sight.  We use a flat sky or small angle approximation.

The correlation between discrete modes is
\begin{align} \label{eq:IIcorrelation}
\langle I_{\bk}I^*_{\bk'}\rangle  = \Omega_s^2 \delta^K_{kk'} P_\bk
\end{align} where $P_{\bk}$ is the discrete power spectrum.  The Kronecker delta implies statistical homogeneity.
The discrete power spectrum is related to the continuous one, $P(k)$, by
\begin{equation}
P_{\bk} = \frac{P(k)}{V_s} = \frac{P(k)}{\Omega_s {\cal D}^2 {\cal L}} = \frac{1}{\Omega_s}C_{\ell,j}.
\end{equation}
Here ${\cal D}$ is the comoving angular size distance to the source volume from the observer.  The angular Fourier coordinate is $\ell = {\cal D} \bk_\perp$ and $j$ denotes the DFT coordinate in the radial direction.  The comoving length of the source volume in the radial direction is $\cal L$.  This serves to define the angular power spectrum in angle and radial coordinates $C_{\ell,j}$.  So in terms of the angular power spectrum
\begin{equation}
\langle I_{\bk}I^*_{\bk'}\rangle = \Omega_s \; C_{\ell,j} \; \delta^{K}_{\ell,\ell'}\delta^{K}_{jj'}.
\end{equation}

Gravitational lensing causes the the observed emission field to be inhomogeneous over a region of the sky with coherent deflection.  The observed intensity after lensing is
\begin{align}
\tilde{I}({ \pmb{\theta}},x_\parallel) & = I\left( \pmb{\theta} - \pmb{\alpha}(\pmb{\theta}), x_\parallel \right) \\
& \simeq  I\left( \pmb{\theta} , x_\parallel \right) - \pmb{\alpha}(\pmb{\theta}) \cdot \nabla_\theta I\left( \pmb{\theta} , x_\parallel \right)  \label{eq:WGLexpansion} \\
& \simeq  I\left( \pmb{\theta} , x_\parallel \right) + \nabla_\theta \Psi(\pmb{\theta}) \cdot \nabla_\theta I\left( \pmb{\theta} , x_\parallel \right) 
\end{align}
where $\pmb{\alpha}(\pmb{\theta})$ is the deflection caused by lensing and $\Psi(\pmb{\theta})$ is the lensing potential.  The deflection field is a potential field to very good approximation.   In Fourier space this becomes
\begin{equation}
\tilde{I}_{\pmb{\ell},j} \simeq I_{\pmb{\ell},j}  +  \frac{1}{\Omega_s} \sum_{\pmb{\ell}'} \pmb{\ell}' \cdot (\pmb{\ell} - \pmb{\ell}')~  I_{\pmb{\ell}',j}~\Psi_{\pmb{\ell} - \pmb{\ell}'}
\end{equation}
From this we can find the correlation between modes to first order
\begin{align} \label{eq:correlation1}
\langle \tilde{I}_{\bell,j}\tilde{I}^*_{\bell-\bL,j'}\rangle  = \Omega_s C_{\ell,j} \delta^{K}_{\bL,0}\delta^K_{jj'} + \delta^K_{jj'}[\bell \cdot \bL C_{\ell,j}+\bL \cdot (\bL-\bl)
C_{|\ell-L|,j}]\Psi_\bL.
\end{align} 

We seek to construct a quadratic estimator of the form
\begin{equation}\label{eq:rawestimator}
\hat{\Psi}(\bL)=\sum_j \sum_{\bl}g(\bl,\bL,j) \; \tilde{I}_{\bell,j}\tilde{I}^*_{\bell-\bL,j}.
\end{equation}
Putting \eqref{eq:correlation1} into \eqref{eq:rawestimator} and requiring the estimator to be unbiased gives the constraint
\begin{equation}
\sum_j \sum_{\bl}g(\bl,\bL,j)[\bell \cdot \bL C_{\ell,j}+\bL \cdot (\bL-\bl)
C_{|\ell-L|,j}]=1.
\label{constraint}
\end{equation}
The variance of this estimator is
\begin{align} \nonumber
&{\cal V} = \sum_j \sum_{j'} \sum_{\bl} \sum_{\bl'} g(\bl,\bL,j) g^*(\bl',\bL,j')\langle I_{\bl,j}I^*_{|\bl-\bL|,j}I^*_{\bl',j'}I_{|\bl'-\bL|,j'}\rangle  \\
&=2\Omega^2_s \sum_j \sum_{\bl}g^2(\bl,\bL,j)C^{\rm tot}_{\ell,j}C^{\rm tot}_{|\bl-\bL|,j},
\end{align} 
where in calculating the fourth order correlations the result for a Gaussian random field has been used and
\begin{equation}
C^{\rm tot}_{\ell,j} = C_{\ell,j} + C^{\rm N}_{\ell},
\end{equation} with $C^{\rm N}_{\ell}$ the thermal noise of the array.

With the standard Lagrangian multiplier technique ${\cal V} $ can be minimized subject to the constraint~\eqref{constraint}.
The result is
\begin{equation} \label{eq:GaussianFilter}
g(\tilde{\bl},\bL,\tilde{j})= A_R \frac{[\tilde{\bell} \cdot \bL C_{\tilde{\ell},\tilde{j}}+\bL \cdot (\bL-\tilde{\bl})
C_{|\tilde{\ell}-L|,\tilde{j}}]}{\Omega^2_sC^{\rm tot}_{\tilde{\ell},\tilde{j}}C^{\rm tot}_{|\tilde{\bl}-\bL|,\tilde{j}}}
~~~~~,~~~~~
A_R =\left[\sum_j \sum_{\bell} \frac{[\bell \cdot \bL C_{\ell,j}+\bL \cdot (\bL-\bl)
C_{|\ell-L|,j}]^2}{\Omega^2_sC^{\rm tot}_{\ell,j}C^{\rm tot}_{|\bl-\bL|,j}} \right]^{-1}.
\end{equation}
The variance of this estimator is 
\begin{equation}
{\cal V} = 2 \left[\sum_j \sum_{\bell}  \frac{[\bell \cdot \bL C_{\ell,j}+\bL \cdot (\bL-\bl)
C_{|\ell-L|,j}]^2}{\Omega^2_sC^{\rm tot}_{\ell,j}C^{\rm tot}_{|\bl-\bL|,j}}\right]^{-1}.
\end{equation}

Now we can verify the agreement of this estimator with previous work by going to continuous Fourier space using 
\begin{equation}
\sum_{\bell} \rightarrow \Omega_s \int \frac{d^2\ell}{(2\pi)^2}
\end{equation} that gives
\begin{equation}
{\cal V} = \Omega_s \left[\sum_j \int \frac{d^2\ell}{(2\pi)^2}  \frac{[\bell \cdot \bL C_{\ell,j}+\bL \cdot (\bL-\bl)
C_{|\ell-L|,j}]^2}{2 C^{\rm tot}_{\ell,j}C^{\rm tot}_{|\bl-\bL|,j}}\right]^{-1},
\end{equation}
which with the identification $\Omega_s \rightarrow (2\pi)^2\delta(0)$ is exactly the \citep{Zahn:2005ap} result.

The behavior of the lensing estimator and noise in the Gaussian approximation has been analyzed in \citep{Zahn:2005ap}.
Higher values of the discretized parallel wave vector $k_\parallel$ (i.e. higher values of $j$) mean higher values of the three-dimensional power spectrum $P(k)$ for each $C_\ell$ value. $P(k)$ is monotonically decreasing on all scales of interest and therefore going to higher values of $j$ the effect of $C_{\ell,j}$ becomes negligible and the signal decays quickly below the thermal noise level. Hence, only a few modes contribute to the estimator \citep{Zahn:2005ap}.

\section{Lensing estimator for unclustered discrete sources}
\label{sec:intens-mapp-lens}

Here we consider a collection of sources with random positions and derive a lensing estimator.  The expansion used in the previous section to describe the action of lensing on the surface brightness (equation~\eqref{eq:WGLexpansion}) is not formally valid here because the deflection angle may be large compared to the size of individual sources.  For this reason we derive the estimator by a different method in this case which also gives some insight into how the estimator works.

The fluctuation in the number of sources with luminosity $L$ in a cell $i$ in real space will be
\begin{equation}
\delta n^L_i = n^L_i - \bar{n}_i
\end{equation}
where the average cell occupation is
\begin{equation}
\bar{n}_i = \bar{\eta}\delta V,
\end{equation}
$\bar{\eta}$ being the average number density and $\delta V$ the volume of a cell.

The surface brightness fluctuation
$\delta S(i)$ can be written 
\begin{equation}
\delta S(i)=\sum_L \delta n^L_i L.
\end{equation}
The continuous limit in luminosity will be taken later.
The discrete Fourier transforming (DFT) of the above expression considering the full 3D case with cells $i=(\ipe,\ipa)$ ($\ipe$ represents the two dimensions perpendicular to the line of sight) is 
\begin{align}
\delta \tilde{S}(j) &= \frac{\Omega_s}{\Npe \Npa}\sum_{\ipa} \sum_{\ipe} \delta S(i)e^{i \frac{2\pi \ipe \jpe}{\Npe}} e^{i \frac{2\pi \ipa \jpa}{\Npa}} \\
&= \frac{\Omega_s}{\Npe \Npa} \sum_{\ipa} \sum_{\ipe} \sum_L L\delta n^L_i e^{i \frac{2\pi \ipe \jpe}{\Npe}} e^{i \frac{2\pi \ipa \jpa}{\Npa}}.
\end{align}
For the average flux in a cell $\bar{S}$ we will have
\begin{equation}
\label{eq:Saver}
\bar{S} = \bar{n}_i \langle L\rangle  = \bar{\eta}\delta V \langle L\rangle .
\end{equation}

Now let us look at correlations between modes:
\begin{align} \nonumber
\langle \delta\tilde{S}(j)\delta \tilde{S}^{*}(j-m)\rangle  
=  \frac{\Omega_s^2}{(\Npe \Npa)^2} \langle \sum_{\ipa} \sum_{\ipe} \sum_L L\delta n^L_i e^{i \frac{2\pi \ipe \jpe}{\Npe}} e^{i \frac{2\pi \ipa \jpa}{\Npa}} 
 \sum_{\ipa'} \sum_{\ipe'} \sum_{L'} L'\delta n^{L'}_{i'} e^{-i \frac{2\pi \ipe' (\jpe-\mpe)}{\Npe}} 
e^{-i \frac{2\pi \ipa' (\jpa-\mpa)}{\Npa}}\rangle .
\end{align}
Since $\langle \delta n^L_i\rangle =0$, only the $i=i', L=L'$ terms contribute:
\begin{align}
\langle \delta\tilde{S}(j)\delta \tilde{S}^{*}(j-m)\rangle  
=  \frac{\Omega_s^2}{(\Npe \Npa)^2} \langle \sum_{\ipa} \sum_{\ipe} \sum_L L^2 (\delta n^L_i)^2 e^{i \frac{2\pi \ipe \mpe}{\Npe}} e^{i \frac{2\pi \ipa \mpa}{\Npa}}\rangle  
\end{align}
The second moment of the number counts in a cell is given by a Poisson distribution $\langle (\delta n_i)^2 \rangle = \overline{n}_i$. If a cell of fixed angular size is magnified by a factor $\mu_i$ the galaxies within it will be a factor $\mu_i$ brighter.  At the same time the true volume of that cell will be a factor of $1/\mu_i$ smaller and so the average number of galaxies will go down by the same factor.  The result is
\begin{align}
\langle \delta\tilde{S}(j)\delta \tilde{S}^{*}(j-m)\rangle  & = 
\frac{\Omega_s^2}{(\Npe \Npa)^2} \langle \sum_{\ipa} \sum_{\ipe} \sum_L \mu_{\ipe} L^2 \bar{n}_i e^{i \frac{2\pi \ipe \mpe}{\Npe}} e^{i \frac{2\pi \ipa \mpa}{\Npa}} \rangle  \label{eq:A9}\\
&=\frac{\Omega_s^2}{(\Npe \Npa)^2} \bar{\eta}\delta V \langle L^2\rangle  \langle \sum_{\ipe} \mu_{\ipe}e^{i \frac{2\pi \ipe \mpe}{\Npe}} \rangle \langle \sum_{\ipa} e^{i \frac{2\pi \ipa \mpa}{\Npa}}\rangle  \\
&= \frac{\Omega_s^2}{\Npe \Npa} \bar{\eta}\delta V \langle L^2\rangle  \tilde{\mu}(\mpe)\delta^K_{\mpa},
\end{align}
where $ \tilde{\mu}(m_\perp)$ is the DFT of the magnification.  

Dividing by the
average flux $\bar{S}$ from Equation (\ref{eq:Saver}) in order to obtain the dimensionless fluctuations, we find
\begin{align} \nonumber
\label{eq:secondmoment}
\langle \Delta\tilde{S}(j) \Delta\tilde{S}^{*}(j-\mpe)\rangle 
&=\frac{\langle \delta\tilde{S}(j) \delta\tilde{S}^{*}(j-\mpe)\rangle }{\bar{S}^2} \\
&=\frac{\Omega_s^2}{\Npe \Npa} \frac{1}{\bar{\eta}\delta V} \frac{\langle L^2\rangle }{\langle L\rangle ^2} \;  \tilde{\mu}(\mpe) \\
&=\frac{\Omega_s^2}{\bar{N}_g}\frac{\langle L^2\rangle }{\langle L\rangle ^2} ~\tilde{\mu}(\mpe) \\  \label{eq:secondmoment2}
&=\Omega_s C^{\rm shot} ~\tilde{\mu}(\mpe),
\end{align}
where we have used $\delta V = \frac{V_{tot}}{\Npe \Npa}$ and $\bar{N}_g=\bar{\eta} V_{tot}$.   \eqref{eq:secondmoment2} defines  $C^{\rm shot}$.  The moments of the luminosity function can now be calculated for a continuous distribution of luminosities.  

We construct the quadratic estimator $\hat{\mu}(\mpe)$ as
\begin{equation}
\hat{\mu}(\mpe)= \frac{1}{\Omega_s C^{\rm shot} }\frac{1}{\Npe \Npa} \sum_{\jpe} \sum_{\jpa} \Delta \tilde{S}(j) \Delta \tilde{S}^{*}(j-\mpe).
\end{equation} 
In this case (with no thermal noise or clustering) the optimal filter ($g(\bell,{\bf L},j)$ in the previous section) is a function of only ${\bf m}_\perp$ and its value is given by the requirement that $\langle \hat{\mu}(\mpe) \rangle = \tilde{\mu}(\mpe) $.
The variance of this estimator is
\begin{align} \nonumber
{\cal V} &= \langle |\hat{\mu(\mpe)}|^2\rangle  = \langle \hat{\mu}(\mpe)\hat{\mu}^{*}(\mpe)\rangle  \\
&= \frac{1}{(\Npe \Npa)^2}  \frac{1}{(\Omega_s C^{\rm shot})^2 }\sum_{\jpe} \sum_{\jpa}  \sum_{\jpe'} \sum_{\jpa'}   \langle  \Delta\tilde{S}(j)\Delta \tilde{S}^{*}(j-\mpe) \Delta \tilde{S}^{*}(j')\Delta \tilde{S}(j'-\mpe) \rangle  \\ \label{eq:variance}
& =\frac{\Omega_s^2}{\bar{N}_g}\frac{\langle L^4\rangle }{\langle L^2\rangle ^2}\left(1+3\frac{\bar{N}_g}{\Npe \Npa}\right) +2\Omega_s^2 \frac{\Npe\Npa-1}{(\Npe \Npa)^2}.
\end{align}

In deriving \eqref{eq:variance} we have used the higher moments of the Poisson number counts in cells.
Note that in the limit $\Npe, \Npa \rightarrow \infty$ only the first term survives.   In the notation of the previous section $\bar{N}_g = \overline{\eta} \Omega_s {\cal D}^2 {\mathcal L}$ where $ \overline{\eta} $ is the density of galaxies.  Note that for 21~cm emission the luminosity is proportional to the HI density so the luminosity moments are also the moments of the HI mass function.

In the weak lensing limit $\mu \simeq 1 + 2\kappa = 1 - \nabla^2\Psi$.  So the estimator for the  lensing potential is related to the estimator for magnification by
\begin{equation}
\hat{\Psi}({\bf L}) =  |{\bf L}|^{-2} \hat{\mu}({\bf L})
\end{equation}
As it turns out, this filter is the same one we would have found by plugging a constant power spectrum into the optimal Gaussian filter~\eqref{eq:GaussianFilter}, but the noise is different.

\section{Lensing estimator for clustered discrete sources}
\label{sec:lens-estim-clust}

In this appendix the results of the two previous appendices are combined to find a lensing estimator and its expected noise in the case of clustered point sources. 
For the $2$-point correlation function we have
\begin{equation}
\langle I(\bx) I(\bx') \rangle  = \frac{1}{\bar{\eta}\delta V}\frac{\langle L^2\rangle }{\langle L\rangle ^2}\delta^K_{\bx\bx'} + \xi_{\bx \bx'}.
\end{equation}
where the first part comes from the Poisson fluctuations in the number counts and the second from the clustering.
In Fourier space
\begin{equation}
\langle I_{\bk} I^*_{\bk'}\rangle  = \Omega_s \; (C_{\ell,j}+C^{\rm shot}) \; \delta^{K}_{\ell,\ell'}\delta^{K}_{jj'}.
\end{equation}
After lensing the observed correlations will be
\begin{align} \nonumber
\langle \tilde{I}_{\bell,j}\tilde{I}^*_{\bell-\bL,j'}\rangle  = \delta^K_{jj'} [\bell \cdot \bL C_{\ell,j}+\bL \cdot (\bL-\bl)
C_{|\ell-L|,j}+L^2 \; C^{\rm shot} ]\; \Psi(\bL)
\end{align}
to first order.  We will not attempt to find an optimal estimator here.  Instead we use an estimator of the form
\begin{equation}
\hat{\Psi}(\bL)= f(\bL) \sum_j \sum_{\bell} \tilde{I}_{\bell,j}\tilde{I}^*_{\bell-\bL,j},
\end{equation} 
where $f(\bL)$ is a normalization. This would be optimal were the sources unclustered.  For an unbiased estimator we require that
$\langle \hat{\Psi}(\bL)\rangle  = \Psi(\bL)$ so we find
\begin{align} \nonumber
f(\bL)&=\bigg\{\sum_j \sum_{\bell}[\bell \cdot \bL C_{\ell,j}+\bL \cdot (\bL-\bl)
C_{|\ell-L|,j}+L^2 \; C^{\rm shot}]\bigg\}^{-1} 
\\ \
&=\bigg\{\sum_j \sum_{\bell}[\bell \cdot \bL C_{\ell,j}+\bL \cdot (\bL-\bl)
C_{|\ell-L|,j}]+(\Npe \Npa)L^2 \; C^{\rm shot}\bigg\}^{-1}.
\end{align}
The variance of the estimator ${\cal V}=\langle \hat{\Psi}(\bL)\hat{\Psi}^*(\bL)\rangle $ is given by
\begin{equation}\label{eq:varianceP+G}
{\cal V}=[f(\bL)]^2 \sum_j \sum_{j'} \sum_{\bell} \sum_{\bell'} \langle I_{\bell,j}I^*_{\bell-\bL,j}I^*_{\bell',j'}I_{\bell'-\bL,j'}\rangle ,
\end{equation}
with
\begin{align} \nonumber
\langle I_{\bell,j}I^*_{\bell-\bL,j}I^*_{\bell',j'}I_{\bell'-\bL,j'}\rangle  =& \Omega_s \frac{1}{\bar{\eta}^3}\frac{1}{(D^2{\cal L})^3}\frac{\langle L^4\rangle }{\langle L\rangle ^4}\left(1+3\frac{\bar{N}_g}{\Npe\Npa}\right) \\ \nonumber
&+ \Omega_s\frac{1}{\bar{\eta}^2}\frac{1}{(D^2{\cal L})^2}\frac{\langle L^3\rangle }{\langle L\rangle ^3}\left(\frac{\Npe \Npa -1}{\Npe \Npa}\right) 
 [C_{|\ell'-L|,j'}+C_{\ell',j'}+C_{|\ell-L|,j}+C_{\ell,j}] \\ \nonumber
&+ \Omega_s\frac{1}{\bar{\eta}^2}\frac{1}{(D^2{\cal L})^2}\frac{\langle L^2\rangle ^2}{\langle L\rangle ^4}\left(\frac{\Npe \Npa -1}{\Npe \Npa}\right) 
 [C_{L,0}+C_{|\ell-\ell'|,|j-j'|}+C_{|\ell+\ell'-L|,j+j'}] \\ \nonumber
&+ \Omega^2_s \frac{1}{\bar{\eta}}\frac{1}{(D^2{\cal L})}\frac{\langle L^2\rangle }{\langle L\rangle ^2}\frac{(\Npe\Npa-1)(\Npe\Npa-2)}{(\Npe\Npa)^2}  \\ \nonumber
& \times [C_{\ell',j'}\delta^K_{\ell,\ell'}\delta^K_{j,j'}\delta^K_{\ell,\ell'}\delta^K_{j,j'}+C_{L-\ell',-j'}
\delta^K_{\ell,L-\ell'}\delta^K_{j,-j'}\delta^K_{\ell,L-\ell'}\delta^K_{j,-j'} \\ \nonumber
&+C_{\ell',j'}\delta^K_{L-\ell,\ell'}\delta^K_{j',-j}\delta^K_{L-\ell,\ell'}\delta^K_{j',-j}
+C_{\ell'-L,j'}\delta^K_{\ell,\ell'}\delta^K_{j,j'}\delta^K_{\ell,\ell'}\delta^K_{j,j'}] \\ \nonumber
&+\Omega^2_s \frac{(\Npe\Npa-1)(\Npe\Npa-2)(\Npe\Npa-3)}{(\Npe\Npa)^3} \\ 
&\times [C_{\ell,j}C_{|\ell'-L|,j'}\delta^K_{\ell,\ell'}\delta^K_{j,j'}\delta^K_{\ell,\ell'}\delta^K_{j,j'}
+C_{\ell,j}C_{\ell',j'}\delta^K_{\ell,L-\ell'}\delta^K_{j,-j'}\delta^K_{\ell,L-\ell'}\delta^K_{j,-j'}].
\label{eq:4thmoment}
\end{align}
Here we have assumed that the clustering and the noise are Gaussian.

We will now calculate the sum over $\bl,\bl',j,j'$ in~\eqref{eq:varianceP+G} separately for each of the terms of the fourth moment (\ref{eq:4thmoment}). We will also 
consider the limit $\Npa \rightarrow \infty$ to simplify the constant factors. Note that, to include the thermal noise of the array in our calculation, we just send $C_{\ell,j} \rightarrow C^{\rm tot}_{\ell,j}$, with
\begin{equation}
C^{\rm tot}_{\ell,j} = C_{\ell,j} + C^N
\end{equation} with $C^N$ the power spectrum of the thermal noise.
 We have
\begin{equation}
{\cal V} = [f(\bL)]^2 \left({\cal I}_0+{\cal I}_1+{\cal I}_2+{\cal I}_3+{\cal I}_4\right).
\end{equation} 
with
\begin{align} \nonumber
{\cal I}_0 &= \sum_j \sum_{j'} \sum_{\bell} \sum_{\bell'} \Omega_s \frac{1}{\bar{\eta}^3}\frac{1}{(D^2{\cal L})^3}\frac{\langle L^4\rangle }{\langle L\rangle ^4} \\ 
&= (\Npe \Npa)^2 \; \Omega_s \frac{1}{\bar{\eta}^3}\frac{1}{(D^2{\cal L})^3}\frac{\langle L^4\rangle }{\langle L\rangle ^4}. \label{eq:I0}
\end{align}
\begin{align} \nonumber
{\cal I}_1 &= \sum_j \sum_{j'} \sum_{\bell} \sum_{\bell'}\Omega_s\frac{1}{\bar{\eta}^2}\frac{1}{(D^2{\cal L})^2}\frac{\langle L^3\rangle }{\langle L\rangle ^3}
 [C^{\rm tot}_{|\ell'-L|,j'}+C^{\rm tot}_{\ell',j'} \\ \nonumber
 &\;\;\;\;\;\;\;\;\;\;\;\;\;\;\;\;\;\;\;\;\;\;\;\;\;\;\;\;\;\;\;\;\;\;\;\;\;\;\;\;\;\;\;\;\;\;\;\;\;\;\;\;\;\;\;\;+C^{\rm tot}_{|\ell-L|,j}+C^{\rm tot}_{\ell,j}]  \\ 
&=\Omega_s\frac{1}{\bar{\eta}^2}\frac{1}{(D^2{\cal L})^2}\frac{\langle L^3\rangle }{\langle L\rangle ^3} \times (\Npe\Npa)\sum_j\sum_{\bl}[2C^{\rm tot}_{\ell,j}+2C^{\rm tot}_{|\ell-L|,j}]. \label{eq:I1}
\end{align}
\begin{align} \nonumber
{\cal I}_2 &= \sum_j \sum_{j'} \sum_{\bell} \sum_{\bell'} \Omega_s\frac{1}{\bar{\eta}^2}\frac{1}{(D^2{\cal L})^2}\frac{\langle L^2\rangle ^2}{\langle L\rangle ^4}
 [C^{\rm tot}_{L,0}+C^{\rm tot}_{|\ell-\ell'|,|j-j'|} \\ \nonumber
& \;\;\;\;\;\;\;\;\;\;\;\;\;\;\;\;\;\;\;\;\;\;\;\;\;\;\;\;\;\;\;\;\;\;\;\;\;\;\;\;\;\;\;\;\;\;\;\;\;\;\;\;\;\;\;\;\;\;\;\;\;\;\;\;\;+C^{\rm tot}_{|\ell+\ell'-L|,j+j'}] \\ \nonumber
&=\Omega_s\frac{1}{\bar{\eta}^2}\frac{1}{(D^2{\cal L})^2}\frac{\langle L^2\rangle ^2}{\langle L\rangle ^4} (\Npe \Npa)^2 C^{\rm tot}_{L,0} \\ 
&+\Omega_s\frac{1}{\bar{\eta}^2}\frac{1}{(D^2{\cal L})^2}\frac{\langle L^2\rangle ^2}{\langle L\rangle ^4}  \sum_j \sum_{j'} \sum_{\bell} \sum_{\bell'} 
[C^{\rm tot}_{|\ell-\ell'|,|j-j'|}+C^{\rm tot}_{|\ell+\ell'-L|,j+j'}]. \label{eq:I2}
\end{align}
\begin{align} \nonumber
{\cal I}_3 &= \sum_j \sum_{j'} \sum_{\bell} \sum_{\bell'} \Omega^2_s \frac{1}{\bar{\eta}}\frac{1}{(D^2{\cal L})}\frac{\langle L^2\rangle }{\langle L\rangle ^2} \\ \nonumber
& \times [C^{\rm tot}_{\ell',j'}\delta^K_{\ell,\ell'}\delta^K_{j,j'}\delta^K_{\ell,\ell'}\delta^K_{j,j'}+C^{\rm tot}_{L-\ell',-j'}
\delta^K_{\ell,L-\ell'}\delta^K_{j,-j'}\delta^K_{\ell,L-\ell'}\delta^K_{j,-j'} \\ \nonumber
&+C^{\rm tot}_{\ell',j'}\delta^K_{L-\ell,\ell'}\delta^K_{j',-j}\delta^K_{L-\ell,\ell'}\delta^K_{j',-j}
+C^{\rm tot}_{\ell'-L,j'}\delta^K_{\ell,\ell'}\delta^K_{j,j'}\delta^K_{\ell,\ell'}\delta^K_{j,j'}] \\
&= \Omega^2_s \frac{1}{\bar{\eta}}\frac{1}{(D^2{\cal L})}\frac{\langle L^2\rangle }{\langle L\rangle ^2} \sum_j \sum_{\bl} [2C^{\rm tot}_{\ell,j}+2C^{\rm tot}_{|\ell-L|,j}]. \label{eq:I3}
\end{align}
\begin{align} \nonumber
{\cal I}_4 &=\sum_j \sum_{j'} \sum_{\bell} \sum_{\bell'} \Omega^2_s [C^{\rm tot}_{\ell,j}C^{\rm tot}_{|\ell'-L|,j'}\delta^K_{\ell,\ell'}\delta^K_{j,j'}\delta^K_{\ell,\ell'}\delta^K_{j,j'} \\ \nonumber
&\;\;\;\;\;\;\;\;\;\;\;\;\;\;\;\;\;\;\;\;\;\;\;\;+C^{\rm tot}_{\ell,j}C^{\rm tot}_{\ell',j'}\delta^K_{\ell,L-\ell'}\delta^K_{j,-j'}\delta^K_{\ell,L-\ell'}\delta^K_{j,-j'}] \\ 
&=\Omega^2_s \sum_j \sum_{\bl} 2C^{\rm tot}_{\ell,j}C^{\rm tot}_{|\ell-L|,j}. \label{eq:I04}
\end{align}
From appendices \ref{app:intens-mapp-lens-continious} and \ref{sec:intens-mapp-lens} we can recognize \eqref{eq:I0} as a pure Poisson clustering term and \eqref{eq:I04} as a pure Gaussian clustering term.  The other terms are from the interplay of these contributions.

To calculate these terms it is convenient to go into continuous $\ell$-space where the integrals can be done numerically.
We have
\begin{align} \nonumber
f(\bL)&=\bigg\{\sum_j \sum_{\bell}[\bell \cdot \bL C_{\ell,j}+\bL \cdot (\bL-\bl)
C_{|\ell-L|,j}+L^2 \; C^{\rm shot}]\bigg\}^{-1}  \\ 
&= \frac{1}{\Omega_s}\bigg\{\sum_j \int \frac{d^2\ell}{(2\pi)^2}[\bell \cdot \bL C_{\ell,j}+\bL \cdot (\bL-\bl)
C_{|\ell-L|,j}+L^2 \; C^{\rm shot}]\bigg\}^{-1}. \label{eq:fbL}
\end{align}
\begin{align} \nonumber
{\cal I}_0 &= \sum_j \sum_{j'} \sum_{\bell} \sum_{\bell'} \Omega_s \frac{1}{\bar{\eta}^3}\frac{1}{(D^2{\cal L})^3}\frac{\langle L^4\rangle }{\langle L\rangle ^4} \\ 
&= (\Npa)^2\Omega^3_s \frac{1}{\bar{\eta}^3}\frac{1}{(D^2{\cal L})^3}\frac{\langle L^4\rangle }{\langle L\rangle ^4}\left(\int\frac{d^2\ell}{(2\pi)^2}\right)^2,  \label{eq:I0cont}
\end{align} where
\begin{equation}
\int d^2\ell = \pi (\ell^2_{\rm max}-\ell^2_{\rm min}).
\end{equation}
\begin{align} \nonumber
{\cal I}_1 =&\Omega_s\frac{1}{\bar{\eta}^2}\frac{1}{(D^2{\cal L})^2}\frac{\langle L^3\rangle }{\langle L\rangle ^3} \times (\Npe\Npa)\sum_j\sum_{\bl}[2C^{\rm tot}_{\ell,j}+2C^{\rm tot}_{|\ell-L|,j}] \\
=&\Omega^3_s\frac{1}{\bar{\eta}^2}\frac{1}{(D^2{\cal L})^2}\frac{\langle L^3\rangle }{\langle L\rangle ^3} \Npa \left(\int\frac{d^2\ell'}{(2\pi)^2}\right)
\sum_j \int\frac{d^2\ell}{(2\pi)^2} [2C^{\rm tot}_{\ell,j}+2C^{\rm tot}_{|\ell-L|,j}].  \label{eq:I1cont}
\end{align}
\begin{align} \nonumber
{\cal I}_2 =& \sum_j \sum_{j'} \sum_{\bell} \sum_{\bell'} \Omega_s\frac{1}{\bar{\eta}^2}\frac{1}{(D^2{\cal L})^2}\frac{\langle L^2\rangle ^2}{\langle L\rangle ^4}
 [C^{\rm tot}_{L,0}+C^{\rm tot}_{|\ell-\ell'|,|j-j'|} \\ \nonumber
& \;\;\;\;\;\;\;\;\;\;\;\;\;\;\;\;\;\;\;\;\;\;\;\;\;\;\;\;\;\;\;\;\;\;\;\;\;\;\;\;\;\;\;\;\;\;\;\;\;\;\;\;\;\;\;\;\;\;\;\;\;\;\;\;\;+C^{\rm tot}_{|\ell+\ell'-L|,j+j'}] \\ \nonumber
&=\Omega^3_s\frac{1}{\bar{\eta}^2}\frac{1}{(D^2{\cal L})^2}\frac{\langle L^2\rangle ^2}{\langle L\rangle ^4} (\Npa)^2 \left(\int\frac{d^2\ell}{(2\pi)^2}\right)^2 C^{\rm tot}_{L,0} \\ \nonumber
&+\Omega^3_s\frac{1}{\bar{\eta}^2}\frac{1}{(D^2{\cal L})^2}\frac{\langle L^2\rangle ^2}{\langle L\rangle ^4}  \sum_j \sum_{j'} \int\frac{d^2\ell}{(2\pi)^2} \int\frac{d^2\ell'}{(2\pi)^2}  \\
& \;\;\;\;\;\;\;\;\;\;\;\;\;\;\;\;\;\;\;\;\;\; \;\;\;\;\;\;\;\;\;\;\;\;\;\;\;\;\; \times [C^{\rm tot}_{|\ell-\ell'|,|j-j'|}+C^{\rm tot}_{|\ell+\ell'-L|,j+j'}].   \label{eq:I2cont}
\end{align}
We are not able to probe the $\kpa=0$ mode (its signal will be removed with foreground cleaning), so we ignore the $C_{L,0}$ contribution to the ${\cal I}_2$ part of the noise and all our plots are with $j_{\rm min}=1$.
\begin{align} \nonumber
{\cal I}_3 &= \Omega^2_s \frac{1}{\bar{\eta}}\frac{1}{(D^2{\cal L})}\frac{\langle L^2\rangle }{\langle L\rangle ^2} \sum_j \sum_{\bl} [2C^{\rm tot}_{\ell,j}+2C^{\rm tot}_{|\ell-L|,j}] \\
&=\Omega^3_s \frac{1}{\bar{\eta}}\frac{1}{(D^2{\cal L})}\frac{\langle L^2\rangle }{\langle L\rangle ^2} \sum_j \int\frac{d^2\ell}{(2\pi)^2} [2C^{\rm tot}_{\ell,j}+2C^{\rm tot}_{|\ell-L|,j}].  \label{eq:I3cont}
\end{align}
\begin{align} \nonumber
{\cal I}_4 &=\Omega^2_s \sum_j \sum_{\bl} 2C^{\rm tot}_{\ell,j}C^{\rm tot}_{|\ell-L|,j} \\ 
&= \Omega^3_s \sum_j   \int\frac{d^2\ell}{(2\pi)^2} 2C^{\rm tot}_{\ell,j}C^{\rm tot}_{|\ell-L|,j}.  \label{eq:I4cont}
\end{align}

Here we should note that foreground subtraction techniques will remove the first few $k_\parallel$ modes \citep{McQuinn:2005hk, Zahn:2005ap}, meaning that we will need to use some $j_{\rm min}>1$ for the lensing reconstruction. Removing only the first few modes would correspond to a relatively simple foreground contamination, while a large number would denote a much more complex problem. Here we should note that in our case the S/N is not greatly affected by removing the few first modes, and we plan to investigate foreground contamination and subtraction techniques in future work.

The behavior of the lensing estimator and noise for clustered discrete sources is more complicated than in the Gaussian approximation due to the complexity of the various contributing terms. As we have stressed in the main text, in our method the Poisson fluctuations contribute both to the signal and the noise of the estimator. More specifically, the contribution of the second moment $C^{\rm shot}$ (see Eq.~\ref{eq:fbL}) is crucial for obtaining a low lensing reconstruction noise level. Of the noise terms the dominant contribution comes from the fourth moment Poisson term ${\cal I}_0$, while the ${\cal I}_2$ term has the smallest contribution. It is also useful to note that, in a similar manner with the Gaussian estimator, the lensing reconstruction noise converges with increasing $j$, so that a relatively small number of modes contributes to the final estimator. For example, our calculations for the source redshift $z=2$ allow us to use a maximum number of parallel modes $j_{\rm max}=63$, but the noise has already converged at $j \sim 40$.  To illustrate that, we show a plot demonstrating the convergence of the lensing reconstruction noise $N(L)$ with $j$ using the SKA2 specifications and Model B for the HI mass function. We see that for this case the noise converges fast (this naturally depends on the interferometer specifications, which determine the contribution of the instrument's thermal noise, as well as the evolution of the HI mass function which affects the signal and the Poisson moments). Also note that the shape of $L^2 N(L)$ is nearly flat up to a scale where noise becomes important --- this behaviour is qualitatively similar to the one of the Gaussian case \citep{Zahn:2005ap}.

\begin{figure}
\centering
\includegraphics[scale=0.4]{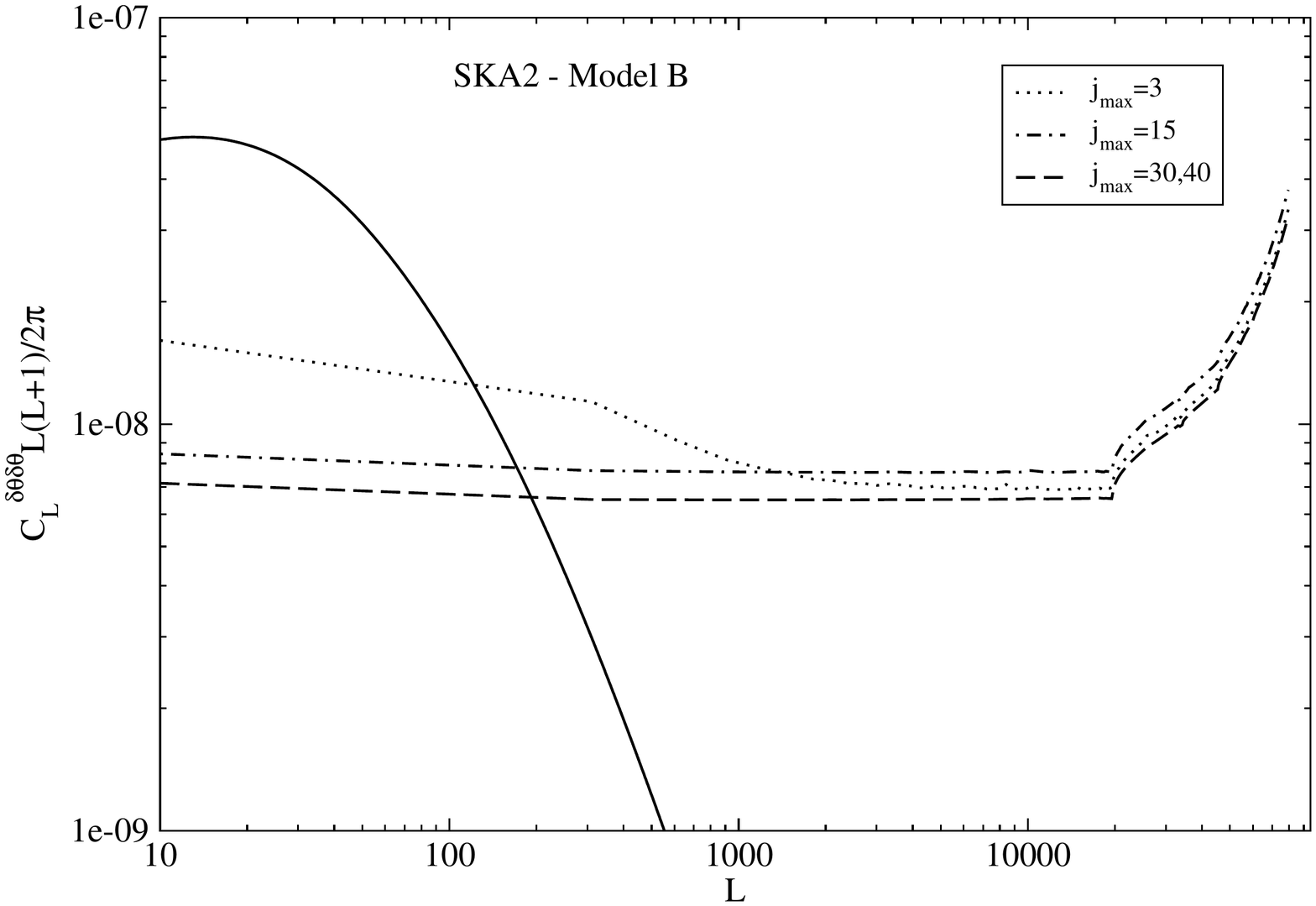}
\caption{Displacement field power spectrum for $z_s=2$  and the corresponding lensing reconstruction noise $N(L)$ for different values of $j_{\rm max}$ using the SKA2 specifications and Model B for the HI mass function.}
\label{fig:CLmodelBjconv}
\end{figure}

\end{document}